\documentclass{article}

\usepackage{microtype}
\usepackage{graphicx}
\usepackage{subfigure}
\usepackage{booktabs} 

\usepackage{hyperref}
\usepackage{amsfonts}
\usepackage{rotating}
\usepackage{booktabs}
\usepackage{balance}
\usepackage{graphicx}
\usepackage{wrapfig}
\usepackage{subfigure}
\usepackage{subcaption}
\usepackage[ruled, vlined, linesnumbered,algo2e]{algorithm2e}
\usepackage{amsmath}
\usepackage{fancyhdr}
\usepackage{bbm}
\usepackage[normalem]{ulem}
\usepackage{enumitem}
\usepackage{multirow}
\usepackage{xcolor}
\usepackage{tikz}
\usepackage{url}
\usepackage{mathtools}
\usepackage{pifont}
\usepackage{makecell}
\usepackage{cleveref}
\usepackage{tabularx}

\setlist{noitemsep, leftmargin=*, topsep=0pt, partopsep=0pt}

\crefformat{section}{\S#2#1#3} 
\crefformat{subsection}{\S#2#1#3}
\crefformat{subsubsection}{\S#2#1#3}

\newcommand{\SYSTEM}{Acela}
\newcommand{\HEU}{Heuristic}
\newcommand{\STRAW}{Strawman}
\newcommand{\NAIVE}{Na\"ive-ML}

\Crefname{figure}{Figure}{Figures}

\DeclareMathOperator*{\argmin}{arg\,min}

\usepackage[accepted]{mlsys2026}

\mlsystitlerunning{Cost-aware Duration Prediction for Software Upgrades in Datacenters}

\begin{document}

\twocolumn[
\mlsystitle{Cost-aware Duration Prediction for Software Upgrades in Datacenters}



\mlsyssetsymbol{equal}{*}

\begin{mlsysauthorlist}
\mlsysauthor{Yi Ding}{to}
\mlsysauthor{Aijia Gao}{goo}
\mlsysauthor{Thibaud Ryden}{goo}
\mlsysauthor{Michal Sedlak}{goo}
\mlsysauthor{Essam Ewaisha}{goo}
\mlsysauthor{Igor Marnat}{goo}
\mlsysauthor{Henry Hoffmann}{ed}
\end{mlsysauthorlist}

\mlsysaffiliation{to}{Elmore Family School of Electrical and Computer Engineering, Purdue University, West Lafayette, IN 47906, USA}
\mlsysaffiliation{goo}{Meta, 1 Meta Way, Menlo Park, CA 94025, USA}
\mlsysaffiliation{ed}{Department of Computer Science, University of Chicago, Chicago, IL 60637, USA}

\mlsyscorrespondingauthor{Yi Ding}{yiding@purdue.edu}

\mlsyskeywords{MLSys, Datacenter, Software Upgrade}

\vskip 0.3in

\begin{abstract}
	

Software upgrades are critical to maintaining server reliability in datacenters. While job duration prediction and scheduling have been extensively studied, the unique challenges posed by software upgrades remain largely under-explored. This paper presents the first in-depth investigation into software upgrade scheduling at datacenter scale. We begin by characterizing various types of upgrades and then frame the scheduling task as a constrained optimization problem. To address this problem, we introduce \SYSTEM{}, a cost-aware duration prediction framework designed to improve upgrade scheduling efficiency and throughput while meeting service-level objectives (SLOs). \SYSTEM{} accounts for asymmetric misprediction costs, strategically selects the best predictive models, and mitigates straggler-induced overestimations. Evaluations on Meta's production datacenter systems demonstrate that \SYSTEM{} significantly increases efficiency of the existing upgrade scheduler by improving upgrade window utilization by 1.25$\times$, increasing the number of scheduled and completed upgrades by 33\% and 41\%, and reducing cancellation rates by 2.4$\times$. The code and data sets will be released after paper acceptance.

\end{abstract}

]



\printAffiliationsAndNotice{}  

\section{Introduction} \label{sec:intro}

Modern internet services operate on hyperscale datacenters housing millions of servers~\cite{govindan2016evolve}. At this scale, any deficiencies in system infrastructure can hurt the datacenter's ability to efficiently handle workloads~\cite{oppenheimer2003internet} and affect billions of users~\cite{naseer2020zero}. To improve system reliability and mitigate the risk of unplanned downtime, regular software upgrade is imperative~\cite{barroso2013datacenter}. A \textbf{software upgrade} refers to the process of replacing an existing version of a software application or system with a newer version that offers enhanced features, improved performance, bug fixes, security patches, and compatibility with updated technologies or hardware. Common software upgrades include upgrades for operating systems, firmware, and kernel~\cite{liu2013zupdate,hu2015explicit,ranganathan2021warehouse}. The key differences between software upgrades and service jobs (e.g., video streaming and search) are detailed in~\cref{sec:diff-sjob}.

Scheduling software upgrades in hyperscale datacenters requires coordination across millions of servers. While job scheduling has been well studied~\cite{krishnaswamy2004estimating,curino2014reservation,boutin2014apollo,jalaparti2015network,jyothi2016morpheus,rajan2016perforator,tumanov2016tetrisched,iorgulescu2017don,chung2018stratus,park20183sigma,jajoo2022case}, software upgrade scheduling at the datacenter scale remains under-explored. To understand its unique challenges, we describe its key steps from Meta's real-world datacenters. In particular, servers are grouped into \textbf{upgrade groups}~\cite{barroso2022datacenter}, which rotate through \textbf{upgrade window} (a fixed time slot for running upgrades). Operators aim to meet service level objectives (SLO), typically requiring 95\% of upgrades to complete within upgrade window~\cite{okuno2019maintenance}. More details of this process are in~\cref{sec:upgrade}.

\begin{figure}[t]
	\centering
	\includegraphics[width=\linewidth]{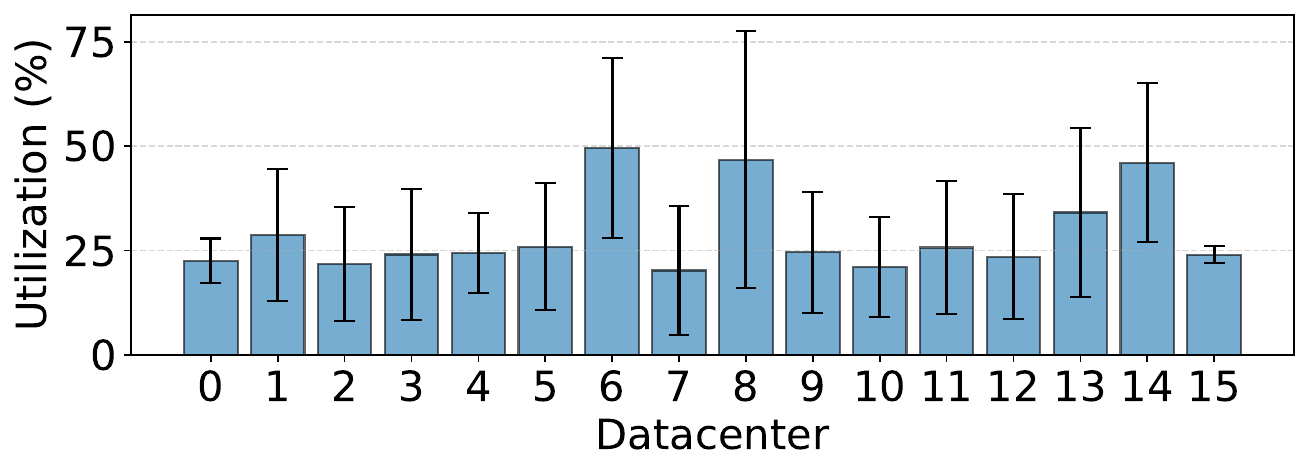} \vspace{-0.1in}
	\caption{Average utilization of upgrade windows for running software upgrades across 16 Meta datacenters over 5 months. The black whiskers indicate standard deviations.} 
	\label{fig:bu}
\end{figure}

To meet SLOs, Meta has adopted a conservative approach that assumes fixed, worst-case durations for all upgrades and schedules few upgrades per server for a long period. While safe, this leads to low efficiency, with frequent idle time and more cycles needed to complete upgrades. As shown in~\Cref{fig:bu}, upgrade window \emph{utilization} (i.e., fraction of time for doing upgrades) across 16 Meta datacenters over 5 months averages just 20–40\%, with some reaching 50\%. High variance further highlights the lack of robust scheduling and underutilization of upgrade windows.


To overcome this limitation, we propose to integrate upgrade duration prediction to improve its scheduling efficiency. We first characterize millions of upgrades from Meta's datacenters to uncover optimization opportunities. We then introduce \SYSTEM{}, a cost-aware duration prediction framework that enhances scheduling throughput and efficiency while meeting SLOs. \SYSTEM{} incorporates three key techniques: (1) it models asymmetric misprediction costs using a tree-based quantile regressor to estimate conditional quantiles, rather than minimizing symmetric loss; (2) it designs a custom scoring function to select models that minimize prediction error while meeting SLOs; and (3) it improves accuracy and mitigates overprediction by training on augmented datasets with stragglers removed.

We implement \SYSTEM{} on top of Meta's existing software upgrade scheduler, integrating duration predictions seamlessly into the scheduling pipeline. We trained \SYSTEM{} on over 4 millions software upgrades collected over three months and tested on 1 million software upgrades during a subsequent month. Our evaluation shows that, compared to the existing scheduler, \SYSTEM{} improves upgrade window utilization by 1.25$\times$ (compared to \Cref{fig:bu}), driven by a 33\% increase in scheduled software upgrades and 41\% more completed software upgrades. Despite handling more software upgrades, \SYSTEM{} meets the SLO by reducing cancellation rates by 2.4$\times$.

To our knowledge, this is the first datacenter-scale study of software upgrades and the first deployed solution to enhance upgrade efficiency. Our key contributions include:
\begin{itemize}
	\item A large-scale characterization and analysis of software upgrades at Meta datacenter production systems.
	\item A formalization of software upgrade scheduling as a constrained optimization problem.
	\item Identification of asymmetric prediction costs unique to upgrade scenarios.
	\item Novel model selection and training methods for upgrade duration prediction.
	\item Deployment of \SYSTEM{} and evaluation on real-world datacenter upgrades with substantial efficiency gains.
\end{itemize}

\section{Background} \label{sec:background}

In this section, we first compare software upgrades with service jobs to highlight their distinct scheduling challenges. We then describe how Meta manages software upgrades, with key terms in \textbf{bold} and defined in~\Cref{tbl:def}.

\begin{table*}[t]
\caption{Explanations of common terms used in the upgrade process in Meta's datacenters.} 
\begin{tabularx}{\textwidth}{l|X} \toprule
\textbf{Term}          & \textbf{Explanation} \\ \midrule	
Software Upgrade   & The process of replacing software with a newer version that improves features, performance, reliability, and compatibility. \\ \hline
Service Job  &  An activity performed on the server (e.g., web search, email, video streaming, etc.) when the server is in normal service operations and not under upgrade or repair. \\ \hline
Upgrade Group & Servers in a rack that share common power and networking switches. \\ \hline
Upgrade Window & A fixed amount of time for an upgrade group to complete all scheduled software upgrades. \\ \hline
Buffer Servers &  Servers that support service job migrated from servers under upgrade.\\ \hline
Overflow Servers & Servers that support service job migrated from servers under repair. \\ \hline
Upgrade Cycle & A period during which all servers undergo software upgrades once. \\ \hline
Upgrade SLO  & At least 95\% of scheduled software upgrades are completed in one upgrade cycle. \\ \hline
Upgrade Efficiency & Number of software upgrades completed within the upgrade window in one upgrade cycle. \\ \hline
Scheduling Throughput & Number of software upgrades scheduled within the upgrade window in one upgrade cycle. \\ \bottomrule  
\end{tabularx}\label{tbl:def}
\end{table*}

\subsection{Software Upgrades vs. Service Jobs}\label{sec:diff-sjob}

\begin{figure}[!t]
	\centering
	\includegraphics[width=\linewidth]{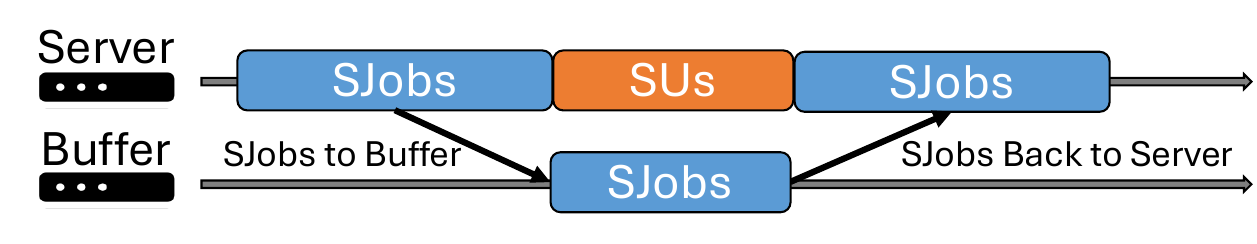} \vspace{-0.15in}
	\caption{Service jobs (SJobs) and software upgrades (SUs) represent distinct phases of a server's lifecycle. } 
	\label{fig:sjob}
\end{figure}

Software upgrades and service jobs represent different phases in a server's lifecycle. Service jobs (e.g., cloud services, search, video streaming) run during normal operation, while software upgrades occur when a server is taken offline and temporarily removed from service. As shown in \Cref{fig:sjob}, these phases do not overlap. During upgrades, service jobs are migrated to a buffer server (see~\cref{sec:upgrade}) and return once the upgrade completes.

\begin{figure}[!t]
	\centering
	\includegraphics[width=0.8\linewidth]{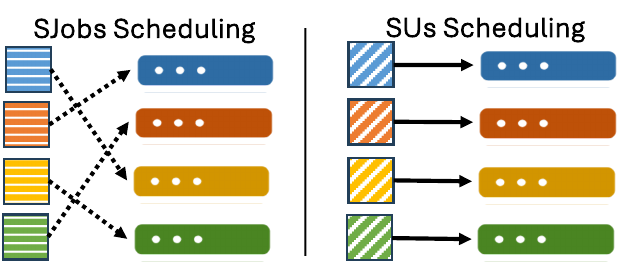} \vspace{-0.1in}
	\caption{Service jobs (SJobs) scheduling vs. software upgrades (SUs) scheduling. Service jobs scheduling is flexible (dotted arrows) with no fixed assignment to servers, while software upgrades scheduling has a fixed assignment (solid arrows), as each server has specific upgrade requirements. } 
	\label{fig:schedule}
\end{figure}

Scheduling software upgrades differs from service job scheduling. As shown in \Cref{fig:schedule}, service jobs are usually flexibly distributed across servers to optimize performance metrics like latency and throughput. In contrast, software upgrades are often tied to specific servers due to their specific upgrade requirements. Moreover, their SLO is more than just latency requirement, but instead requires that at least 95\% of upgrades complete within a designated upgrade window. This combination of tied server assignments and unique SLO makes upgrade scheduling challenging.

\subsection{Circle-based Upgrade Process in Meta}\label{sec:upgrade}

\begin{figure}[!t]
	\centering
	\includegraphics[width=0.8\linewidth]{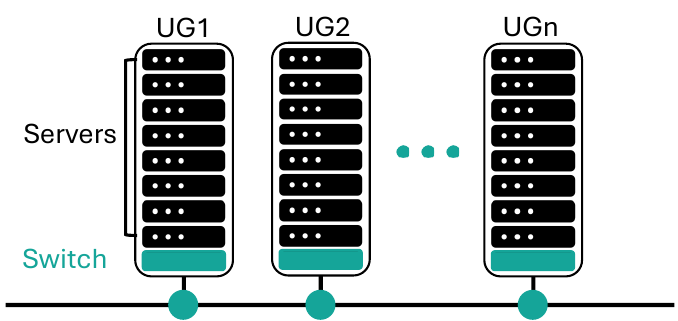} \vspace{-0.1in}
	\caption{Upgrade groups (UGs). Each UG consists of multiple servers in a rack that share power supply and networking switch. All UGs are connected to the internet.} 
	\label{fig:UG}
\end{figure}

In Meta's datacenters, servers are organized into racks, with each rack sharing power and networking infrastructure. As shown in \Cref{fig:UG}, all servers within a rack form an \textbf{upgrade group (UG)}—the basic unit for managing upgrades, which proceed cyclically at the UG level.

At any time, a server can play one of three roles: (1) a \textbf{regular server} running service jobs or software upgrades, (2) a \textbf{buffer servers} hosting service jobs migrated from servers undergoing upgrades, or (3) an \textbf{overflow servers} handling jobs from failed servers undergoing repair~\footnote{Upgraded servers return on schedule, while failed servers have uncertain recovery times.}. These roles rotate to balance service load and upgrade progress.

\begin{figure}[!t]
	\centering
	\includegraphics[width=0.7\linewidth]{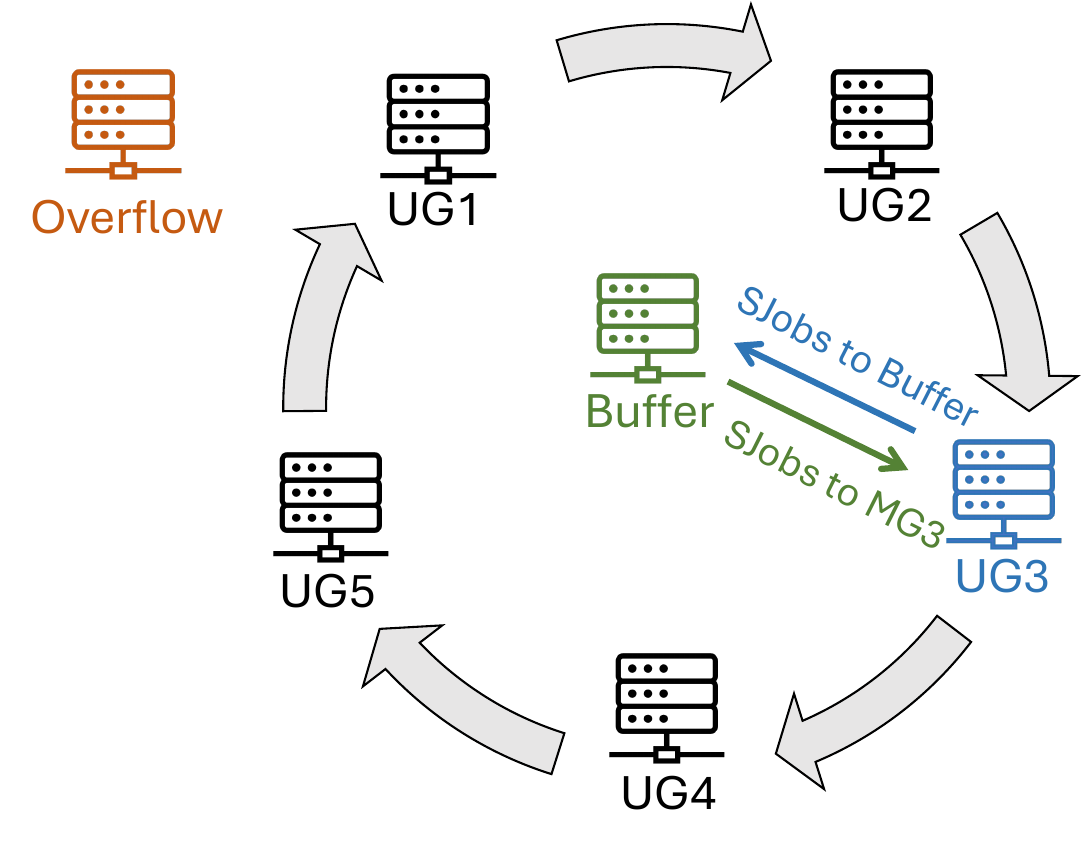}  \vspace{-0.1in}
	\caption{An illustration of an upgrade cycle consisting of 5 regular UGs, 1 buffer UG, and 1 overflow UG. In this cycle, UG1-5 are regular servers handling service job or software upgrades. When upgrade is scheduled for UG3, it will migrate all its service job to the Buffer UG before running software upgrades. After the upgrade window closes, service job will be returned to UG3. Then, upgrade will move to UG4.} 
	\label{fig:cycle}
\end{figure}

As illustrated in \Cref{fig:cycle}, most servers run service jobs or upgrades, while a small subset serves as buffer or overflow. During an upgrade cycle, when a UG is scheduled, its service jobs are temporarily migrated to buffer servers. After the upgrade window ends, servers reboot and service jobs return, with the next UG entering the upgrade phase. If a server takes too long to run software upgrades and misses its upgrade window, it is marked for repair and its jobs are migrated to overflow servers to maintain service continuity.

Within a UG, software upgrades are performed in a partially sequential manner; no more than a fixed fraction (e.g., 25\%) of servers can be upgraded simultaneously due to limited buffer capacity for migrated service jobs. Each server scheduled for upgrade receives at least one upgrade. To meet the SLO, it is essential to avoid overloading a server with too many upgrades in a single cycle, which would increase the risk of exceeding the upgrade window otherwise. If a server fails to finish on time, it is marked for repair, and its service job is moved to an overflow server to maintain service continuity. Crucially, missing the SLO makes it difficult to distinguish between slow and faulty servers without costly manual inspection. Meeting such SLO will reduce these inspections, improve operational efficiency, and help identify true hardware issues early.

\section{Problem Analysis}

We first discuss the key challenges in duration prediction and then analyze how different types of prediction errors affect scheduling outcomes.

\begin{figure*}[!t]
	\centering
	\includegraphics[width=\linewidth]{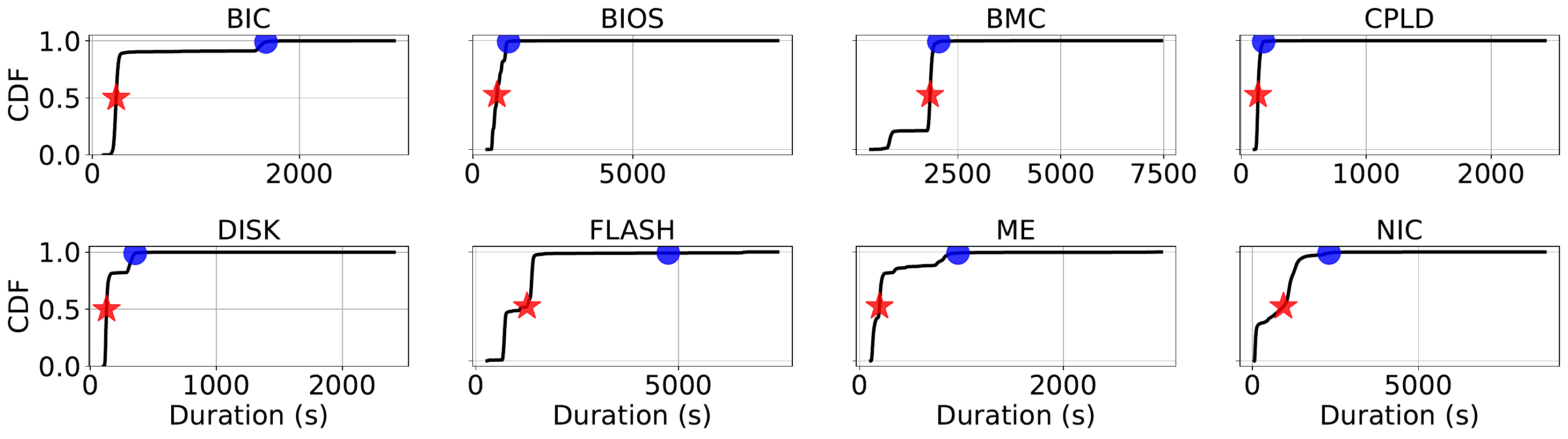} \vspace{-0.15in}
	\caption{CDFs of firmware upgrades durations. Median and p99 are marked by red stars and blue dots.}
	\label{fig:cdf}
\end{figure*}

\subsection{Challenges in Prediction}

As shown in \Cref{fig:bu}, the current upgrade scheduler exhibits low \textbf{upgrade efficiency}, resulting in long idle time within upgrade windows and requiring more cycles to complete all upgrades. This inefficiency is from assuming uniform, worst-case durations across all upgrades, ignoring their type or hardware context. To improve upgrade efficiency, we must accurately predict durations for different upgrades. However, there are two challenges in obtaining high-quality duration predictions, outlined below.

\textbf{Challenge \#1.} Software upgrades include various types, each with different duration distributions. For instance, firmware upgrades---key software upgrades in
datacenters~\cite{liu2013zupdate,hu2015explicit,ranganathan2021warehouse} and the focus of our evaluation (more details in~\cref{sec:evaluation})---include types such as BIC~\cite{bic}, BIOS~\cite{bios}, CPLD~\cite{cpld}, DISK~\cite{disk}, FLASH~\cite{flash}, ME~\cite{me}, NIC~\cite{nic}, and BMC~\cite{bmc}. \Cref{fig:cdf} shows the cumulative distribution functions of these upgrade durations. The diversity and long tails in distributions make prediction challenging.

\textbf{Challenge \#2.} Our ultimate goal is to maximize upgrade scheduling efficiency while meeting SLOs, whereas most machine learning predictors aim to maximize accuracy. This creates a mismatch between system objectives and prediction accuracy, so we should design our predictor to directly support the system goal rather than accuracy alone.

\subsection{Four Types of Mispredictions}\label{sec:mispredict} 

To design the upgrade duration predictor, we analyze four misprediction types, each affecting outcomes differently.

\textbf{Underprediction} occurs when predicted durations are shorter than actual ones, potentially leading to overloaded upgrade windows. This can delay server recovery and stall the upgrade cycle.

\textbf{Extreme underprediction} significantly underestimates durations, often due to stragglers caused by hardware failures. These require repair and overflow server use, and if present in training data, can bias models toward extreme overprediction.

\textbf{Overprediction} occurs when predicted durations are longer than actual ones, causing the scheduler to underutilize the upgrade window. However, mild overprediction is generally preferable, as it aligns with the scheduling objective of ensuring timely completion.

\textbf{Extreme overprediction} significantly overestimates durations, mirroring the existing worst-case-based scheduler and resulting in low upgrade efficiency.

In summary, effective predictors should slightly overpredict to meet the SLO, minimize straggler influence during training, and account for variation across upgrade types.

\section{\SYSTEM{}} \label{sec:framework}

We present \SYSTEM{}, a cost-aware duration prediction framework that improves scheduling throughput and upgrade efficiency while meeting the SLO. This section walks through step-by-step development of \SYSTEM{}.

\subsection{Design Principles}

\SYSTEM{} is guided by the following design principles:

\begin{itemize}
	\item \textbf{Account for asymmetric costs.} As discussed in~\cref{sec:mispredict}, underprediction hurts SLO compliance more than overprediction. Therefore, the training loss function should reflect such imbalance.
	\item \textbf{Balance upgrade efficiency and SLO compliance.} Overprediction helps meet SLOs by preventing upgrade overload, but excessive overprediction can reduce upgrade efficiency.
	\item \textbf{Limit extreme overprediction.} Stragglers in training data can cause significant overprediction. Reducing their influence is key to improving prediction quality.
	\item \textbf{Capture duration variability.} Different upgrade types have different duration distributions (\Cref{fig:cdf}). The models should account for these differences.
\end{itemize}

Next, we will present the techniques we developed in \SYSTEM{} following the principles above.

\subsection{Quantile Regression for Asymmetric Costs}\label{sec:qr}

\SYSTEM{} uses quantile gradient boosting trees to account for asymmetric costs between under- and overpredictions in software upgrade duration predictor. We will first explain conditional quantiles, then quantile loss, and finally how quantile gradient boosting trees work.

\textbf{Conditional quantile.} Most regression methods predict conditional mean by minimizing a symmetric loss function. Let $Y$ represent software upgrade duration and $X$ the software upgrade features. Given a $X=x$, the conditional mean minimizes the expected squared error loss (also known as the L2 loss function):

\begin{align*}
E(Y|X=x) = \argmin_z E\{(Y-z)^2|X=x \}.
\end{align*}

However, this loss function is symmetric and cannot address asymmetric costs in software upgrade duration prediction. To overcome this limitation, \SYSTEM{} estimates the conditional quantile rather than the conditional mean. Let $F(y|X=x)=P(Y\leq y|X=x)$ be the conditional distribution function for $Y$. The $\tau$-quantile $q_{\tau}(x)$ of $Y$ is defined such that the probability of $Y$ being smaller than $q_{\tau}(x)$ is, for a given $X=x$, exactly equal to $\tau$:

\begin{align*}
q_{\tau}(x) = \inf \{y: F(y|X=x) \geq \tau \}.
\end{align*}

\begin{figure}[!t]
	\begin{center}
			\includegraphics[width=0.95\linewidth]{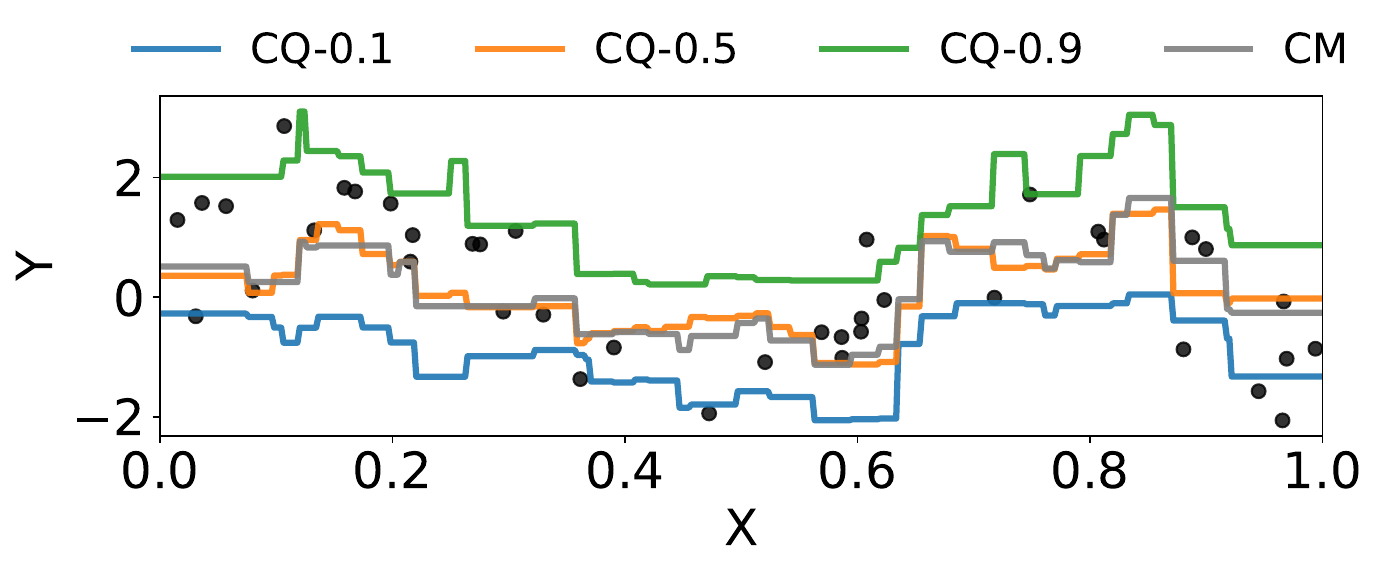} 
		\end{center} 
	\caption{An example for comparing conditional quantiles to conditional mean. \texttt{CQ-}$\tau$ represents conditional quantile when $\tau=0.1, 0.5, 0.9$, respectively. \texttt{CM} represents conditional mean. The black dots are true data.}
	\label{fig:quantile}
\end{figure}


The conditional quantile gives information about the distribution of software upgrade duration. \Cref{fig:quantile} compares conditional quantile to conditional mean when fitting the same dataset. As seen, \texttt{CQ-}0.9 is higher than \texttt{CQ-}0.5, \texttt{CQ-}0.1, and \texttt{CM}, reflecting that for the same $x$, \texttt{CQ-}0.9 predicts a longer software upgrade duration. By selecting a higher value for $\tau$ (e.g., $>0.5$), we can overpredict software upgrade durations.

\textbf{Quantile loss.} To estimate conditional quantile from data, we need to minimize a quantile loss function. The quantile loss function $L_{\tau}$ ($0<\tau<1$) is defined as

\begin{align}\label{eq:quantile}
L_{\tau}(y, \hat{y}) = 
\tau(y-\hat{y})\mathbb{I}(y\geq\hat{y}) + (\tau-1)(y-\hat{y})\mathbb{I}(y<\hat{y}),
\end{align}

\noindent where $\tau$ is the input quantile parameter, $\mathbb{I}(\cdot)=1$ if its $\cdot$ is true and 0 otherwise, $y$ and $\hat{y}$ are true and predicted durations, respectively. When quantile regression underpredicts,  $\mathbb{I}(y\geq \hat{y})=1$, and the loss function becomes $\tau(y-\hat{y})$. When quantile regression overpredicts, $\mathbb{I}(y< \hat{y})=1$, and the loss function becomes $(\tau-1)(y-\hat{y})$. 

The quantile loss penalize underpredictions ($y\geq \hat{y}$) and overpredictions ($y<\hat{y}$) asymmetrically. When $\tau>0.5$ it penalizes more on underprediction than overprediction. When $\tau<0.5$, it penalizes more on overprediction than underprediction. When $\tau=0.5$, it penalizes them equally. Therefore, to overpredict software upgrade durations, $\tau$ should be set above 0.5. 


\textbf{Quantile Gradient boosting trees (QGBT).} \SYSTEM{} uses QGBT to estimate conditional quantiles~\cite{meinshausen2006quantile}. QGBT is a tree-based ensemble method that builds trees sequentially, with each new tree trained to correct the residuals of the previous one. Trees are added until a set limit is reached or no further improvement is observed. The final prediction is the sum of all tree outputs. The key difference between QGBT and standard GBT lies in how data is handled during tree growth. Standard GBT retains only the mean of the data in each node after splitting, while QGBT preserves the full distribution of values. This enables QGBT to model conditional distributions and adjust predictions based on the desired quantile, allowing fine control over under- or overprediction.

When using QGBT for training predictors, \SYSTEM{} needs to pre-specify the quantile parameter $\tau$. Choosing this parameter is essential to optimize upgrade efficiency while meeting the SLO, leading us to the next subsection.

\subsection{Custom Scoring for Model Selection}\label{sec:score}

The optimal quantile parameter $\tau$ in QGBT should balance upgrade efficiency and SLO compliance. To find the best $\tau$, \SYSTEM{} introduces a custom score function for model selection. It trains multiple QGBT models across different quantiles and evaluates them on a validation set designed to reflect real-world deployment—containing upgrades that occur chronologically after the training set. The model with the lowest score is selected. The custom score function is:

\begin{align}\label{eq:score}
\texttt{score} = \begin{cases}
		\texttt{MAE}       & \text{if }~\texttt{OPR} \geq \texttt{SLO} \\
		(\alpha\cdot(1-\texttt{OPR})+1)\cdot \texttt{MAE}  &  \text{if }~\texttt{OPR} <\texttt{SLO}
\end{cases}
\end{align}

\texttt{MAE} is the mean absolute error on the validation set, measuring prediction accuracy (lower is better). \texttt{OPR} is the overprediction rate, which is the fraction of upgrades with predicted durations exceeding actual durations. The SLO is set at 95\%, meaning at least 95\% of upgrades must complete within the upgrade window. The parameter $\alpha$ controls the penalty for underprediction.

The score function is designed with the following intuition:
If \texttt{OPR} $\geq$ SLO, it suggests the SLO will be met, so the focus shifts to minimizing \texttt{MAE} to improve scheduling throughput by avoiding excessive overprediction. If \texttt{OPR} $<$ SLO, it implies a risk of missing the SLO, and a penalty is applied and increases more severe as \texttt{OPR} drops further below the target. This scoring mechanism balances SLO compliance with upgrading efficiency.

\begin{figure}[!t]
	\centering
	\includegraphics[width=0.9\linewidth]{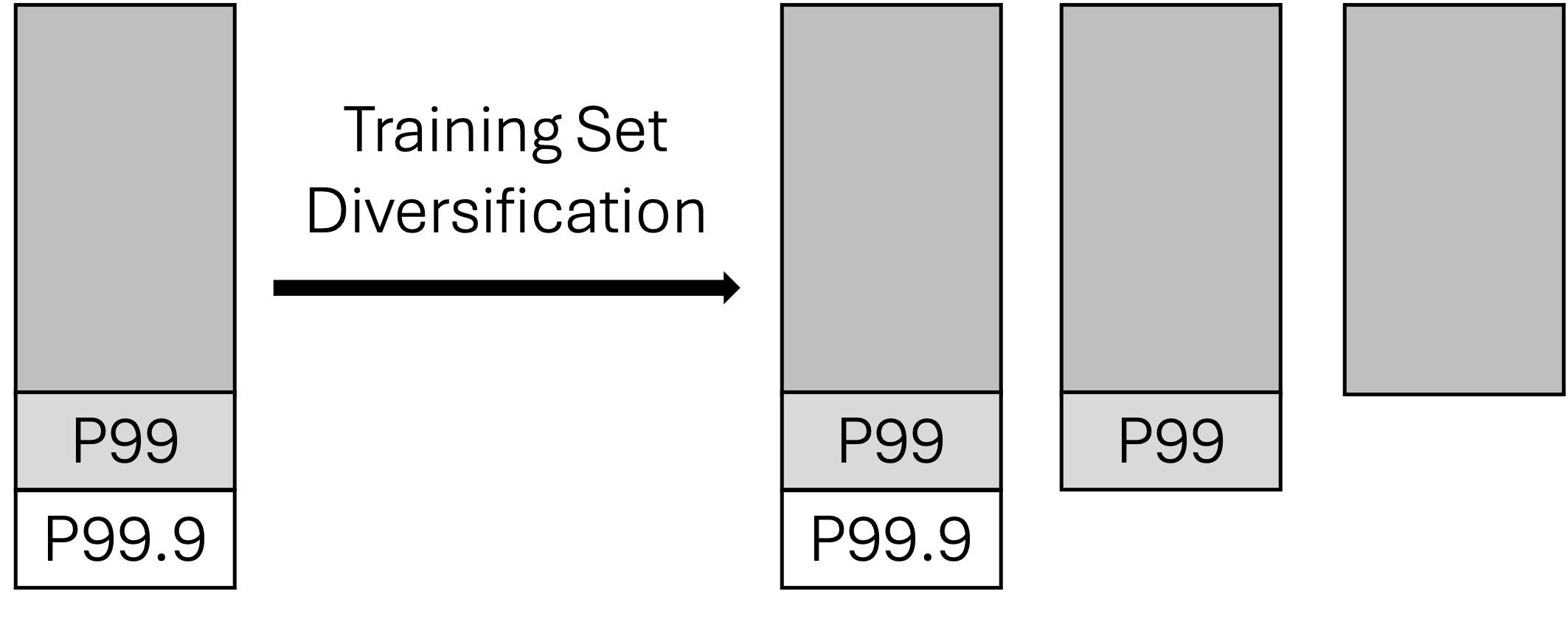}
	\caption{An illustration of training set diversification. } 
	\label{fig:div} 
\end{figure}

\subsection{Training Set Diversification for Reducing Bias}\label{sec:clean}

The quality of prediction results depends not only on the model but also on the data. As shown in \Cref{fig:cdf}, stragglers exist in the training data, which can cause extreme overprediction, decreasing upgrade efficiency.

To address this, \SYSTEM{} diversifies the training dataset to reduce prediction bias from stragglers. The idea is to create several truncated datasets by excluding software upgrades with extreme long durations, such as those in the 99th and 99.9th percentiles (p99 and p99.9), as shown in~\Cref{fig:div}. For each truncated dataset, \SYSTEM{} trains several QGBT models using different quantile parameters and then selects the best-performing model with the lowest score (\Cref{eq:score}), evaluated with the validation set. The empirical evaluation indicates that this approach ensures more accurate and reliable predictions (\Cref{tbl:straggler}).

\subsection{Putting It All Together}\label{sec:together}

\SYSTEM{} trains software upgrade duration predictors through a multi-step process. It builds a separate QGBT model for each upgrade type to capture duration variability. In an online setting, it continuously updates training and validation data by adding recent upgrades and discarding outdated ones. It enriches the dataset by creating truncated datasets that remove stragglers (those above p99 and p99.9) and trains QGBT models on multiple quantiles for each dataset. Using the SLO and penalty parameter $\alpha$, \SYSTEM{} scores all models and selects the one with the lowest value from Equation~\ref{eq:score}. It then deploys the best-performing model for each upgrade type to make predictions.

\section{Integration with Scheduling}\label{sec:deploy}


\begin{figure}[t]
	\centering
	\includegraphics[width=0.95\linewidth]{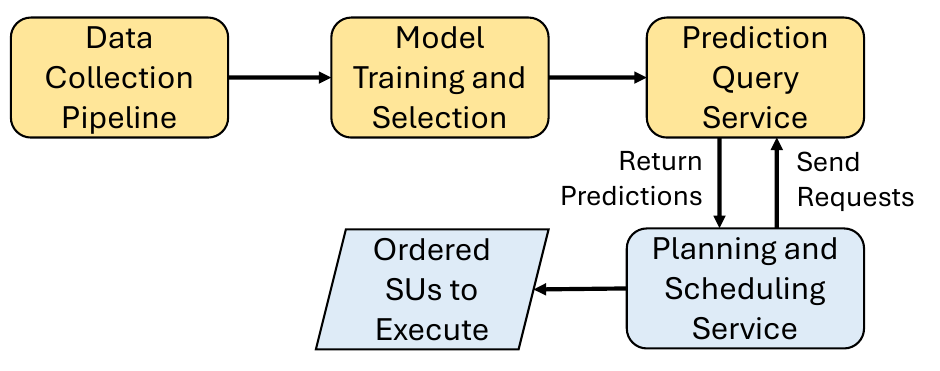} 
	\caption{Integrating \SYSTEM{} into scheduling. The modules of the existing software upgrade scheduler are in blue. \SYSTEM{}'s new modules are in yellow. } 
	\label{fig:workflow} 
\end{figure}

\SYSTEM{} is integrated into existing datacenter software upgrade schedulers as an enhancement layer rather than a replacement. We first describe how \SYSTEM{} interfaces with the scheduler through additional system modules, and then explain the upgrade scheduling logic driven by priority and duration awareness. Importantly, the improvement in upgrade efficiency comes from the joint integration of duration prediction and scheduling decisions, rather than from either component in isolation.

\textbf{Integration.} \Cref{fig:workflow} shows the workflow of integrating \SYSTEM{} into the scheduling process, where three new modules are added to the existing scheduler: the Data Collection Pipeline (\textbf{DCP}), the Model Training and Selection (\textbf{MTS}), and the Prediction Query Service (\textbf{PQS}). All modules communicate via custom remote procedure calls (RPCs) and RESTful APIs~\cite{fielding2000architectural}. Next, we describe each module in detail.

The DCP module continuously logs software upgrade data. These features include software upgrade details like type, upgrade versions (current and future), hardware specifics (CPU architecture, core count, RAM, and disk size), and manufacturing and upgrade information, such as device specifics, location, and last upgrade date. Details of features are in the Appendix.

The MTS module uses this data to train and select QGBT models. It splits the data into training and validation sets, ensuring the validation set has newer software upgrades to handle changes in data distribution. It uses the LightGBM library~\cite{ke2017lightgbm} to build QGBT models, tuning hyperparameters like the number of trees, tree depth, and learning rate. The MTS adjusts the quantile parameter using the custom score function and regularly updates models with new data while removing old data.

The PQS module receives the selected model from MTS to provide duration predictions for specific software upgrades requested by the scheduling service. It processes JSON-formatted requests with server ID, software upgrade type, and desired upgrade version, returning duration predictions to the scheduling service, which then generates an ordered list of software upgrades for scheduling.


\textbf{Scheduling Logic.}
We extend the existing upgrade scheduler with duration-aware decision making while preserving its original priority semantics. An upgrade’s priority mainly depends on three factors: (1) \emph{Upgrade history}: an upgrade type receives higher priority if it has not been performed recently. (2) \emph{Dependency constraints}: if upgrade A depends on upgrade B, then B is ranked higher and scheduled first; if A is subsumed within B, we prioritize B and avoid redundant scheduling of A. (3) \emph{Stakeholder interests}: upgrades deemed critical by operators or service owners may be promoted regardless of historical frequency.

\SYSTEM{} leverages duration prediction to facilitate scheduling under these priorities. First, when two upgrades have equal priority, the scheduler favors the one with the shorter predicted duration to improve overall throughput. Second, when priority and predicted duration conflict (e.g., upgrade A has higher priority but a longer duration than B), the scheduler typically executes the higher-priority upgrade first to maintain policy compliance. For complex or ambiguous scenarios, datacenter operators retain the ability to intervene and make the final decision.

\subsection{Discussion}

We conducted failure mode analysis by focusing on how prediction and system integration can break the upgrade SLO under operational constraints like rack-based upgrade groups, limited buffer capacity, and repair/overflow handling. Specifically, we analyzed (1) correlated underprediction within an upgrade group (UG): even small errors become harmful if many servers in the same rack are underestimated together; (2) distribution shift after new upgrade versions, since \SYSTEM{} retrains weekly and depends on upgrade-version features; (3) straggler removal side effects; and (4) service-level dependency failures such as Prediction Query Service timeouts or missing features, which could silently fall back to unsafe defaults. Each failure mode can be tied to measurable signals they already report (cancellation rate, utilization, OPR) and tested via replay or small fault injections, making it practical to execute. 

For worst-case analysis, since upgrades are scheduled in fixed windows and failure to finish may trigger repair workflows, the worst case is when a subset of upgrades run much longer than expected and push the upgrade group over the window, causing cancellations and operational ambiguity between slow and faulty servers. This scenario highlights that robustness is not only about prediction accuracy, but also about bounding tail execution time and providing operators with clear signals to distinguish genuine failures from duration outliers. We leave the in-depth worse-cases analysis for future work.
\section{Evaluation}\label{sec:evaluation}

In this section, we present our evaluation results, covering methodology, main production results, and detailed predictions across upgrade types and design choices.

\subsection{Methodology}

\textbf{Evaluation setup.} We train \SYSTEM{} on over 4 million software upgrades collected over three months and evaluate it on 198 UGs, spanning nearly 1 million upgrades from a month of real datacenter operations. Each UG schedules at least 1,000 upgrades to ensure robust analysis. Among these, 151 UGs use \SYSTEM{}, while 47 follow the Heuristic baseline with uniform worst-case durations; because \SYSTEM{} increases scheduling volume, it is applied to more UGs. The SLO requires at least 95\% of upgrades to finish within the upgrade window, and we set the score penalty parameter $\alpha$ to 10. 
We chose this value through cross-validation over $[0.1, 1, 10, 100, 1000]$. In practice, we revisit it every month to keep up with the upgrade evolution.
To limit straggler effects, we truncate training data at the 99th and 99.9th percentiles and retrain \SYSTEM{} weekly as new data arrives. 
In this work, we focus on firmware upgrades because they represent the most operationally challenging upgrade category in production datacenters. Empirically, firmware upgrades exhibit both long execution times (typically 1–2 hours) and substantial duration variance across servers and upgrade groups, making them difficult to schedule efficiently and reliably under fixed maintenance windows. In contrast, other common upgrade types show different characteristics. OS upgrades have comparable durations (1–2 hours) but are significantly more stable and predictable due to standardized workflows and mature tooling. Kernel and network switch upgrades are generally much shorter (often under 30 minutes), which limits their impact on scheduling efficiency and reduces the need for duration prediction. As a result, firmware upgrades present the highest uncertainty and operational difficulty, making them the most suitable and impactful target for duration prediction techniques such as \SYSTEM{}. \Cref{tbl:fw} lists eight firmware upgrades studied in this paper, which together represent over 99\% of all firmware upgrades.

\begin{table}[!t]
\footnotesize
	\caption{Eight software upgrades evaluated in this paper.} 
	\begin{tabularx}{\linewidth}{l|X} \toprule
		\textbf{SU}          & \textbf{Explanation} \\ \midrule	
		BIC& The bridge interconnect upgrade~\cite{bic}. \\ \hline
		BIOS& The basic input/output system upgrade~\cite{bios}. \\ \hline
		BMC& The board management controller upgrade~\cite{bmc}. \\ \hline
		CPLD& The complex programmable logic device upgrade for power management~\cite{cpld}. \\ \hline
		 DISK& The storage disk upgrade~\cite{disk}. \\ \hline
		FLASH& The solid state drive upgrade~\cite{flash}. \\ \hline
		ME& The Intel management engine upgrade~\cite{me}.  \\ \hline
		NIC& The network interface card upgrade~\cite{nic}. \\ \bottomrule  
	\end{tabularx}\label{tbl:fw}
\end{table}

\textbf{Comparisons.} In real-world evaluation, we compare \SYSTEM{} against \textbf{\HEU}, which applies fixed heuristics and conservative duration estimates long used by Meta's datacenter schedulers. We further evaluate \SYSTEM{} through simulation, comparing it with \textbf{\STRAW}, which predicts durations using per-type training averages, and \textbf{\NAIVE}, a standard ML model optimized for accuracy. We cannot deploy \STRAW{} or \NAIVE{} in production because each server experiences a specific upgrade only once, and deploying them solely for testing would require substantial implementation and validation efforts. Our simulation closely mirrors real-world behavior, with less than 1\% deviation in testing. For fairness, \NAIVE{} uses the same gradient boosting tree model as \SYSTEM{}, but predicts conditional means instead of conditional quantiles.

\textbf{Metrics.} We measure upgrade efficiency using two metrics: \textbf{utilization}, the fraction of upgrade window time spent actively upgrading (higher is better), and \textbf{cancellation rate}, the fraction of upgrades that fail to complete within the window (lower is better). We also report the total number of scheduled and completed upgrades.

\subsection{Main Production Results}\label{sec:res-main}

\begin{figure}[!t]
	\centering
	\includegraphics[width=\linewidth]{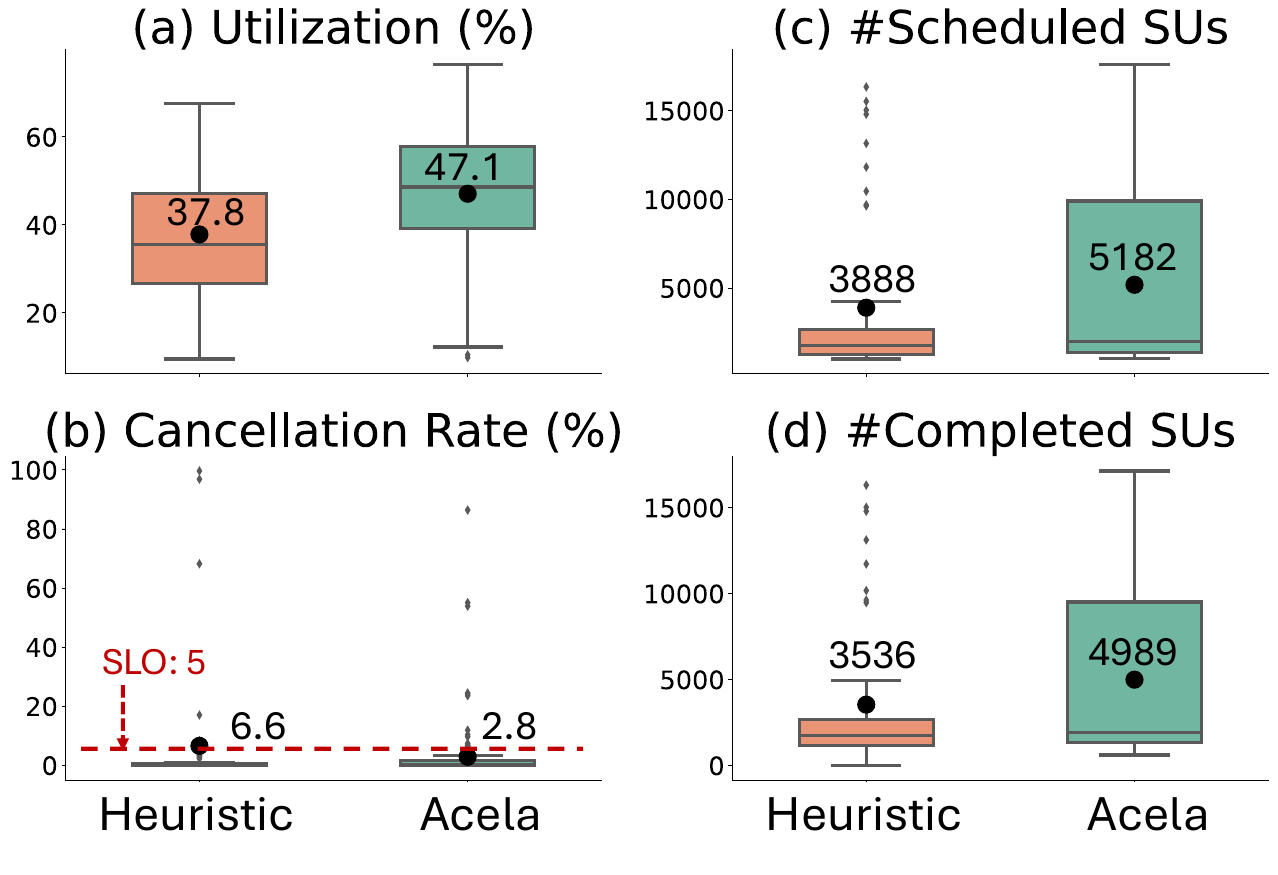} 
	\caption{Main production results between \HEU{} and \SYSTEM{} (\cref{sec:res-main}). The black dot and the number show the mean of the data displayed in the boxplot. The SLO of 95\% software upgrades completed requires the cancellation rate below 5\%. On average, \SYSTEM{} meets it and \HEU{} fails to do so. } 
	\label{fig:res-main}
\end{figure}

For real-world evaluation, we compare upgrade window utilization, cancellation rate, number of scheduled software upgrades, and number of completed software upgrades between \HEU{} and \SYSTEM{} in~\Cref{fig:res-main}. On average, \SYSTEM{} achieves 1.25$\times$ higher upgrade window utilization than \HEU{}, driven by a 33\% increase in scheduled software upgrades and 41\% more completed software upgrades. This improvement stems from \SYSTEM{}'s ability to predict varying durations for different software upgrades, rather than assuming uniform, long durations for all software upgrades as \HEU{}. Despite handling more software upgrades, \SYSTEM{} meets the SLO by reducing the software upgrade cancellation rate by 2.4$\times$. \HEU{} fails to meet the SLO, which requires a cancellation rate below 5\%. This is because \HEU{}, which does not account for software upgrade durations, may schedule actual long-duration software upgrades in a upgrade window. In contrast, \SYSTEM{} prevents this by incorporating duration predictions into its scheduling decisions.

\subsection{Simulation Results on an Upgrade Group}\label{sec:res-sim}

\begin{table}[!t]
	\caption{Simulation results on a UG (\cref{sec:res-sim}). \#Scheduled and \#Completed are the number of scheduled and completed software upgrades respectively, and a higher value is better. CR refers to cancellation rate, and a lower value is better. The best number in each metric is in \textbf{bold}.}
	\begin{center}
\footnotesize
	\begin{tabular}{lllll}\toprule
		& \HEU{} & \STRAW{} & \NAIVE{}    & \SYSTEM{}  \\ \midrule
		\#Scheduled & 9241      & 11604    & \textbf{12041}  & 11972  \\ \hline
		\#Completed & 9234      & 11495    & 11941  & \textbf{11963}  \\ \hline
		CR          & 0.08\%    & 0.94\%   & 0.83\% & \textbf{0.08\%} \\ \bottomrule
	\end{tabular}\label{tbl:res-sim}
	\end{center}
\end{table}

To compare \SYSTEM{} against other baseline predictors, we simulate scheduling process on a real-world UG with 14,515 total requested software upgrades. \Cref{tbl:res-sim} presents the results for number of scheduled software upgrades, number of completed software upgrades, and cancellation rate (CR). \HEU{} schedules and completes the fewest software upgrades due to its assumption of fixed, long durations for all software upgrades. In contrast, \STRAW{}, \NAIVE{}, and \SYSTEM{} schedule and complete up to 30\% more software upgrades than \HEU{}. While \NAIVE{} schedules the most jobs, \SYSTEM{} completes the highest number of software upgrades. Despite completing the most, \SYSTEM{} maintains a low cancellation rate, matching \HEU{} and performing 11.75$\times$ better than \STRAW{} and 10.4$\times$ better than \NAIVE{}.

\SYSTEM{} completes the most software upgrades with the lowest cancellation rate by effectively integrating duration prediction and priority. Although \NAIVE{} schedules more software upgrades, it overloads servers, resulting in a high cancellation rate. \STRAW{} faces similar issues. \SYSTEM{} avoids this by using a unique training process that slightly overpredicts software upgrade durations, allowing it to schedule effectively, complete more jobs, and keep the cancellation rate low.

\begin{figure}[t]
	\centering
	\begin{subfigure}
		\centering       
		\includegraphics[width=\linewidth]{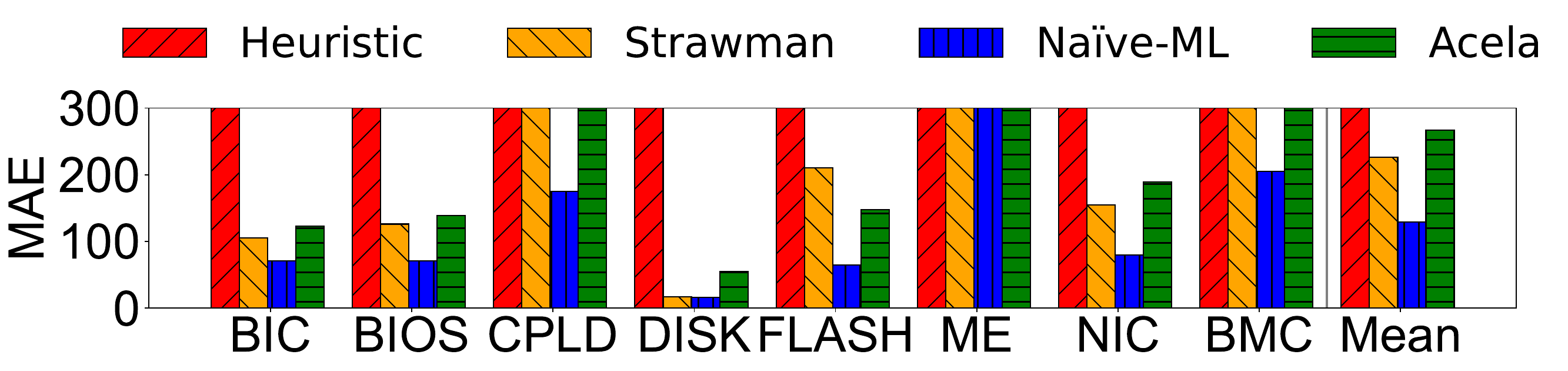} 
	\end{subfigure} 
	\begin{subfigure}
		\centering 
		\includegraphics[width=\linewidth]{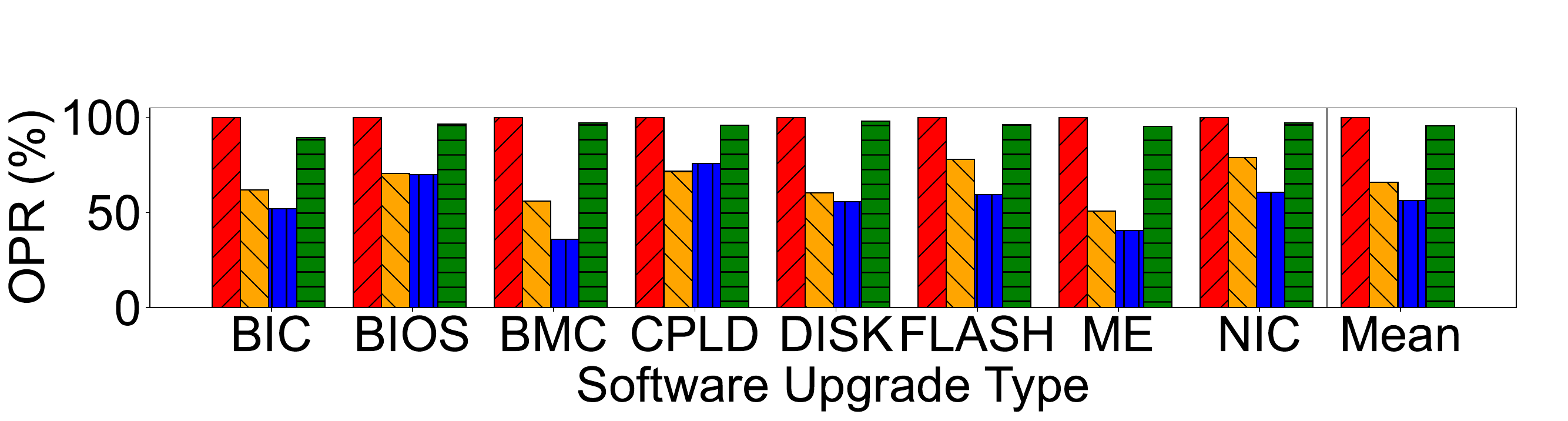} 
	\end{subfigure} 
	\caption{Prediction accuracy (MAE) and overprediction rate (OPR) for each firmware upgrade type (\cref{sec:res-pred}). To improve readability, MAE bar charts are capped at 300. Lower MAE indicates higher prediction accuracy. }\label{fig:res-pred}
\end{figure}

\subsection{Duration Prediction Results}\label{sec:res-pred}

We show duration prediction results for different firmware upgrade types, evaluated using two metrics: \emph{prediction accuracy (MAE)} and \emph{overprediction rate (OPR)}. MAE is the mean absolute error between predicted and actual durations, and a lower value indicates better accuracy. OPR measures the fraction of software upgrades that are overpredicted in an UG, which directly affects the ability to meet the SLO. An OPR slighter above the SLO target (i.e., 95\%) indicates near-optimal results.

\Cref{fig:res-pred} show two rows of bar charts for MAE and OPR respectively, where the x-axis represents the firmware upgrade, y-axis represents MAE/OPR, and the last column Mean is the arithmetic mean over all firmware upgrade types. For MAE, \HEU{} performs the worst, with MAE values 38-79$\times$ higher than the other three due to its assumption of fixed and long durations for all software upgrades, regardless of type. \NAIVE{} achieves the best MAE, with values 1.8-79$\times$ lower than the other three, as it uses a na\"ive machine learning technique---training gradient boosting trees with the squared loss function---focused on optimizing prediction accuracy. \STRAW{} and \SYSTEM{} fall in between, with \STRAW{} using average durations per software upgrade type, while \SYSTEM{} optimizes a custom score function that balances SLO and upgrade efficiency.

\begin{figure}[t]
	\centering
	\includegraphics[width=0.95\linewidth]{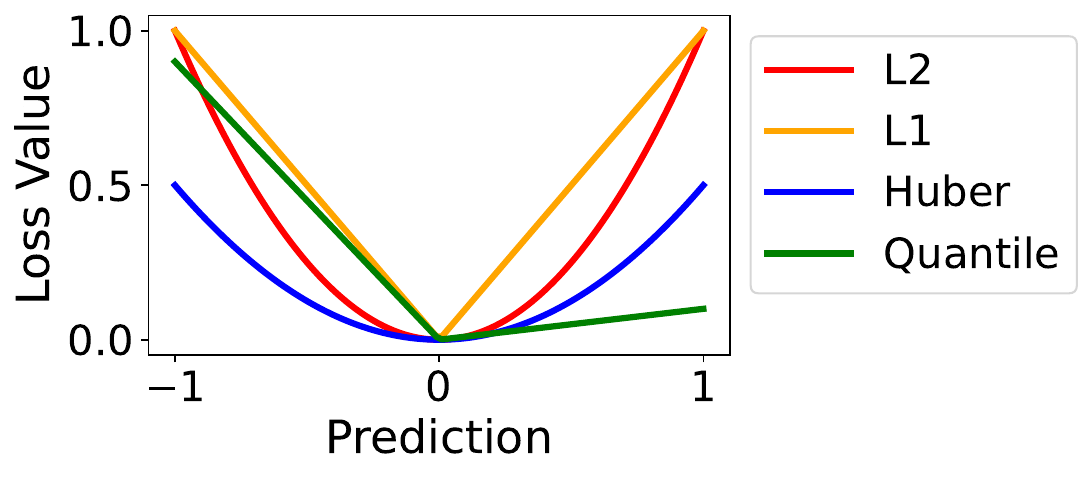}  
	\caption{\textbf{Comparing different loss functions.} $\delta$ is set to 1 in the \texttt{Huber} loss. $\tau$ is set 0.9 in the \texttt{Quantile} loss. } 
	\label{fig:loss}
\end{figure}

\begin{figure}[t]
	\centering
	\begin{subfigure}
		\centering
		\includegraphics[width=\linewidth]{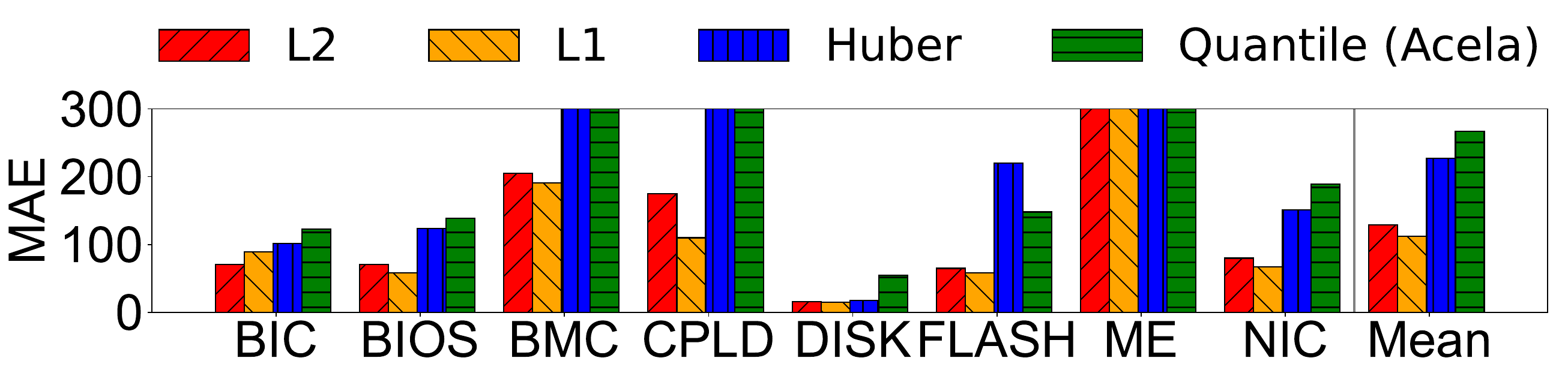} 
	\end{subfigure} 
	\begin{subfigure}
		\centering 
		\includegraphics[width=\linewidth]{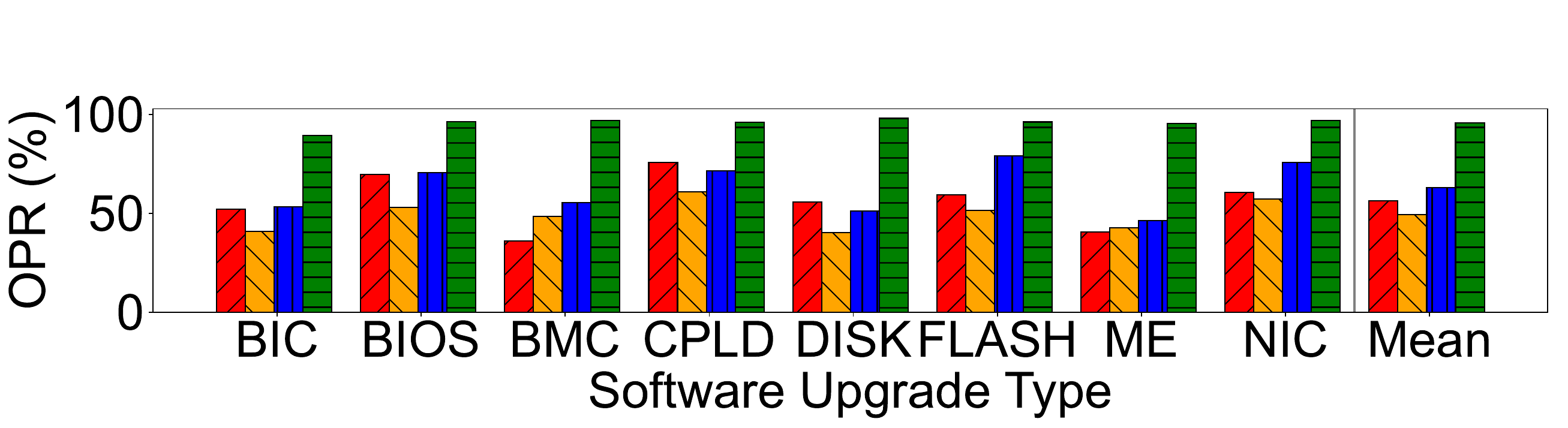} 
	\end{subfigure} 
	\caption{Prediction accuracy (MAE) and overprediction rate (OPR) for different loss functions (\cref{sec:res-loss}). \texttt{L2}, \texttt{L1}, and \texttt{Huber} are symmetric loss functions, while \texttt{quantile (\SYSTEM{})} is asymmetric. To improve readability, MAE bar charts are capped at 300, with mean values labeled directly. Lower MAE indicates higher prediction accuracy. }\label{fig:res-loss}
\end{figure}

\subsection{Versus Other Loss Functions}\label{sec:res-loss}

This section justifies our choice of loss function in \SYSTEM{} by comparing the quantile loss—an asymmetric loss function—to three commonly used symmetric regression losses: \texttt{L2}~\cite{gareth2013introduction}, \texttt{L1}~\cite{tibshirani1996regression}, and \texttt{Huber}~\cite{meyer2021alternative}.
As shown in \Cref{fig:loss}, where the x-axis indicates the prediction and the y-axis shows the corresponding loss, all three symmetric losses (\texttt{L2}, \texttt{L1}, and \texttt{Huber}) exhibit balanced penalty structures for under- and overpredictions. In contrast, the quantile loss introduces asymmetry—when the quantile parameter is set to 0.9 (i.e., above the median), it penalizes underpredictions more heavily than overpredictions.

\begin{figure}[t]
	\centering
	\begin{subfigure}
		\centering
		\includegraphics[width=\linewidth]{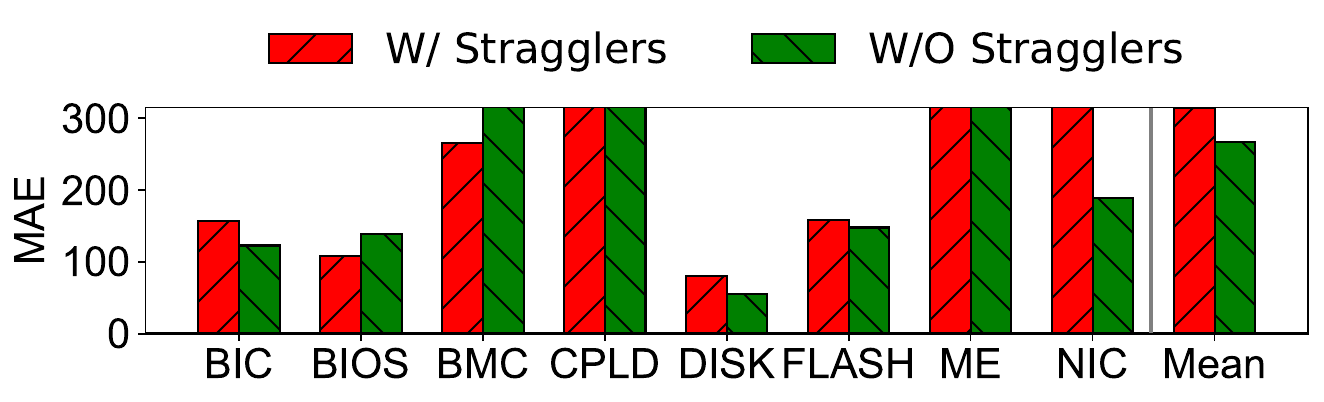} 
	\end{subfigure} 
	\begin{subfigure}
		\centering 
		\includegraphics[width=\linewidth]{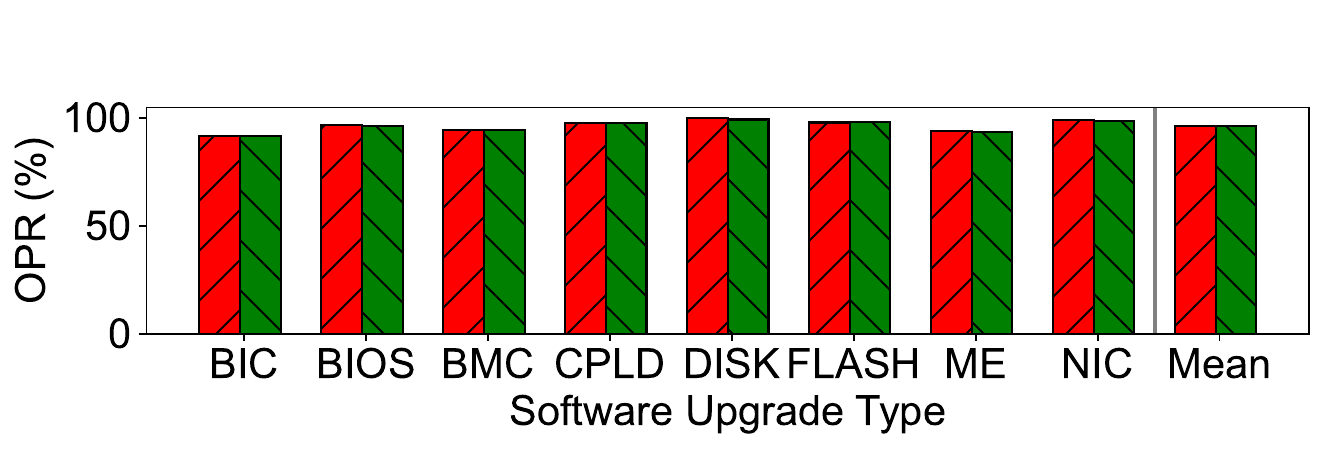} 
	\end{subfigure}
	\caption{Prediction accuracy (MAE) and overprediction rate (OPR) with vs. without stragglers (\cref{sec:res-straggler}). To improve readability, MAE bar charts are capped at 300, with mean values labeled directly. Lower MAE reflects better prediction accuracy. Given the 95\% SLO, an OPR slightly above 95\% is considered nearly-optimal. While both \texttt{W/ Straggler} and \texttt{W/O Stragglers} achieve similar OPR, the latter has better accuracy. }\label{fig:res-straggler}
\end{figure}

\begin{table*}[t]
	\caption{Prediction accuracy (MAE), overprediction rate (OPR), and score (\Cref{eq:score}) across quantiles and straggler removal levels. \texttt{q\_X\_Y} denotes training at \texttt{X}\% quantile with \texttt{Y}-th percentile stragglers removed. \textbf{Bold} models are selected by \SYSTEM{}.}
	\small
	\begin{center}
		\begin{tabular}{l|lll|lll|lll|lll} \toprule
			& \multicolumn{3}{c}{\textbf{BIC}}   & \multicolumn{3}{c}{\textbf{BIOS}} & \multicolumn{3}{c}{\textbf{BMC}} & \multicolumn{3}{c}{\textbf{CPLD}}    \\ \midrule	
			Model                & MAE    & OPR    & Score   & MAE    & OPR   & Score   & MAE    & OPR   & Score   & MAE     & OPR    & Score    \\ \hline
			\SYSTEM{}   & 33     & 95\%     & 33      & 85     & 98\%    & 85  & 179     & 99\%     & 179    & 334    & 99\%    & 334         \\ \hline
			\texttt{q\_95\_99.0} & \textbf{33} & \textbf{95\%} & \textbf{33} & \textbf{85}  & \textbf{98\%} & \textbf{85} & 125          & 94\%           & 75159      & \textbf{334} & \textbf{99\%}  & \textbf{334}  \\ \hline
			\texttt{q\_95\_99.9} & 36     & 96\%     & 36      & 87     & 98\%    & 87  & 148     & 94\%     & 89158      & 845    & 99\%    & 845         \\ \hline
			\texttt{q\_95\_100}  & 37     & 96\%     & 37      & 88     & 98\%    & 88   & 149     & 94\%     & 89843      & 1876   & 99\%    & 1876       \\ \hline
			\texttt{q\_99\_99.0} & 85     & 99\%     & 85      & 145    & 99\%    & 145  & \textbf{179} & \textbf{99\%}  & \textbf{179}   & 2004   & 99\%    & 2004      \\ \hline
			\texttt{q\_99\_99.9} & 160    & 99\%     & 160     & 284    & 99\%    & 284  & 210     & 99\%     & 210      & 2341   & 99\%    & 2341       \\ \hline
			\texttt{q\_99\_100}  & 214    & 99\%     & 214     & 348   & 99\%  & 348   & 211  	   & 99\%    & 211     & 2660   & 99\%    & 2660      \\ \bottomrule
			& \multicolumn{3}{c}{\textbf{DISK}} & \multicolumn{3}{c}{\textbf{FLASH}}   & \multicolumn{3}{c}{\textbf{ME}}  & \multicolumn{3}{c}{\textbf{NIC}} \\ \midrule	
			Model                & MAE    & OPR    & Score   & MAE    & OPR   & Score   & MAE    & OPR   & Score   & MAE     & OPR    & Score    \\  \hline
			\SYSTEM{}   & 46      & 100\%    & 46        & 39     & 97\%     & 39      & 424    & 96\%    & 424     & 196    & 100\%   & 196           \\ \hline    
			\texttt{q\_95\_99.0} & \textbf{46}  & \textbf{100\%} & \textbf{46} & \textbf{39} & \textbf{97\%} & \textbf{39} & \textbf{424} & \textbf{96\%} & \textbf{424} & 82           & 94\%           & 49051            \\ \hline
			\texttt{q\_95\_99.9}  & 82      & 100\%    & 82  & 40     & 97\%     & 40      & 424    & 96\%    & 424     & 90     & 94\%    & 54235     \\ \hline
			\texttt{q\_95\_100}   & 89      & 100\%    & 89   & 40     & 97\%     & 40      & 426    & 96\%    & 426     & 93     & 94\%    & 55847    \\ \hline     
			\texttt{q\_99\_99.0} & 25      & 94\%     & 14835  & 214         & 100\%         & 214         & 838          & 99\%          & 838          & \textbf{196} & \textbf{100\%} & \textbf{196}  \\ \hline
			\texttt{q\_99\_99.9} & 25      & 94\%     & 15272 & 231    & 100\%    & 231     & 861    & 99\%    & 861     & 321    & 100\%   & 321         \\ \hline
			\texttt{q\_99\_100}  & 23      & 93\%     & 16108  & 231    & 100\%    & 231     & 964    & 99\%    & 964     & 322    & 100\%   & 322     		
			\\ \bottomrule 
		\end{tabular}\label{tbl:straggler}
	\end{center}
\end{table*}

\Cref{fig:res-pred} shows duration prediction results using each loss function. We can see that \SYSTEM{} has the highest MAE, performing 1.2-2.4$\times$ worse in prediction accuracy compared to the other three loss functions. It is because all three loss functions try to minimize the differences between true and predicted durations, while \SYSTEM{} optimizes a designed score function rather than the prediction accuracy.

However, prediction accuracy alone does not provide the full picture---it is equally important to assess whether the SLO is met. When looking at OPR, all methods using symmetric loss functions yield low OPRs, around 50\%, which falls short of the targeted SLO. This happens because they estimate the conditional mean, while \SYSTEM{} estimates a conditional quantile that encourages overprediction. As a result, only \SYSTEM{} with the quantile loss meets the SLO, achieving an OPR of 95\%.

\subsection{Impact of Removing Stragglers}\label{sec:res-straggler}

We analyze the impact of removing stragglers—identified by their high tail latency at p99 and p99.9—from the training set. \Cref{fig:res-straggler} compares prediction performance with stragglers included (\texttt{W/ Stragglers}) versus excluded (\texttt{W/O Stragglers}, as used by \SYSTEM{}). The OPRs are nearly identical: 96.275\% vs. 96.075\%, since including stragglers only slightly increases overprediction by biasing predictions upward. However, \texttt{W/ Stragglers} shows 1.2$\times$ worse MAE, confirming that removing stragglers improves prediction accuracy and, in turn, upgrade efficiency.

We further explore the combined effects of quantile selection, straggler removal, and score-based model selection. \Cref{tbl:straggler} reports MAE, OPR, and scores across models trained at 95\% and 99\% quantiles, with straggler removal at the 99th, 99.9th, and 100th percentiles (no removal). Two key insights emerge. First, all models chosen by \SYSTEM{} (lowest scores) remove stragglers, confirming the value of straggler filtering. Second, models meeting the 95\% SLO consistently achieve lower scores, while those falling short are heavily penalized, reflecting the design of the score function.

\begin{figure}[!t]
	\centering
		\includegraphics[width=\linewidth]{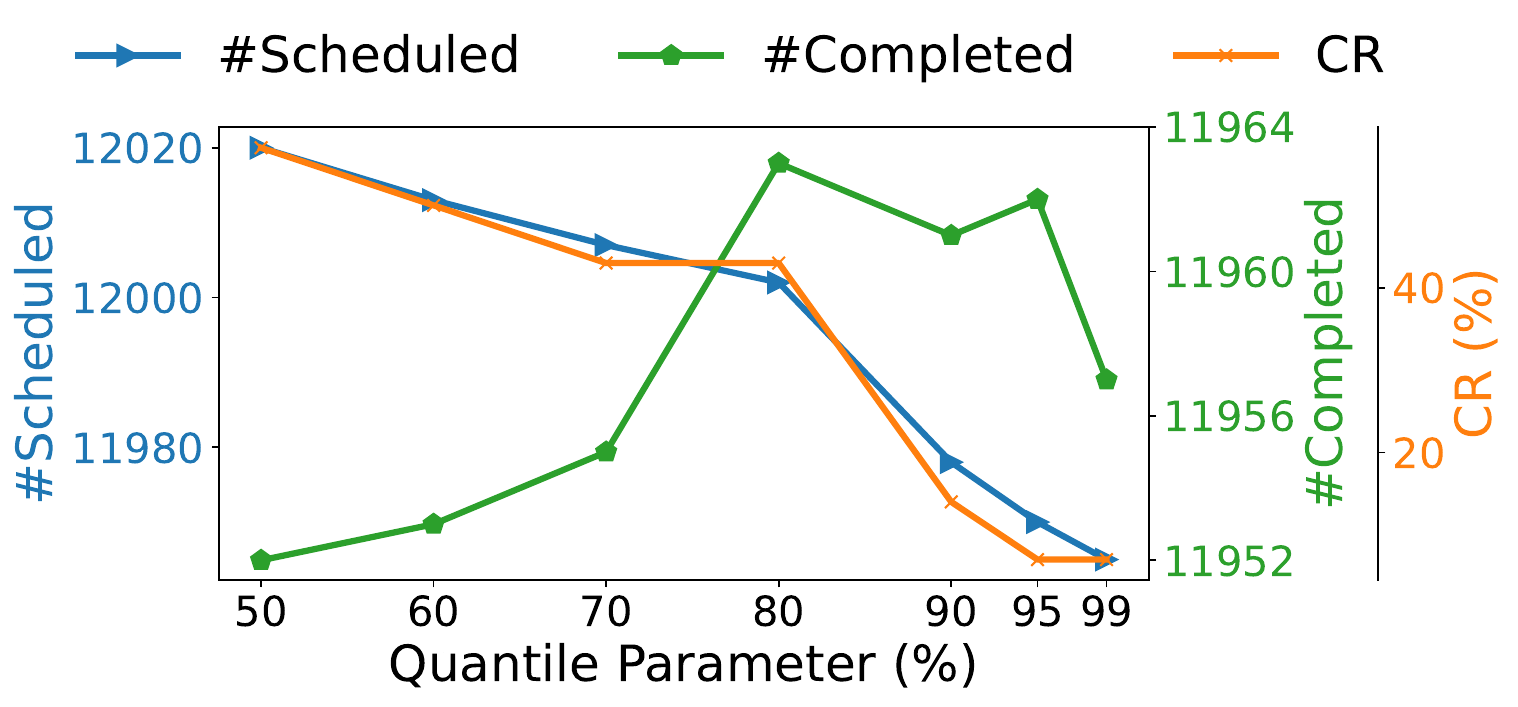} 
	\caption{Sensitivity analysis of quantile parameter $\tau$ on the upgrade outcomes in the number of scheduled software upgrades (\#Scheduled), the number of completed software upgrades (\#Completed), and cancellation rate (CR) (\cref{sec:res-sen}).}\label{fig:res-sen2}
\end{figure}

\begin{figure}[!t]
	\centering
	\includegraphics[width=\linewidth]{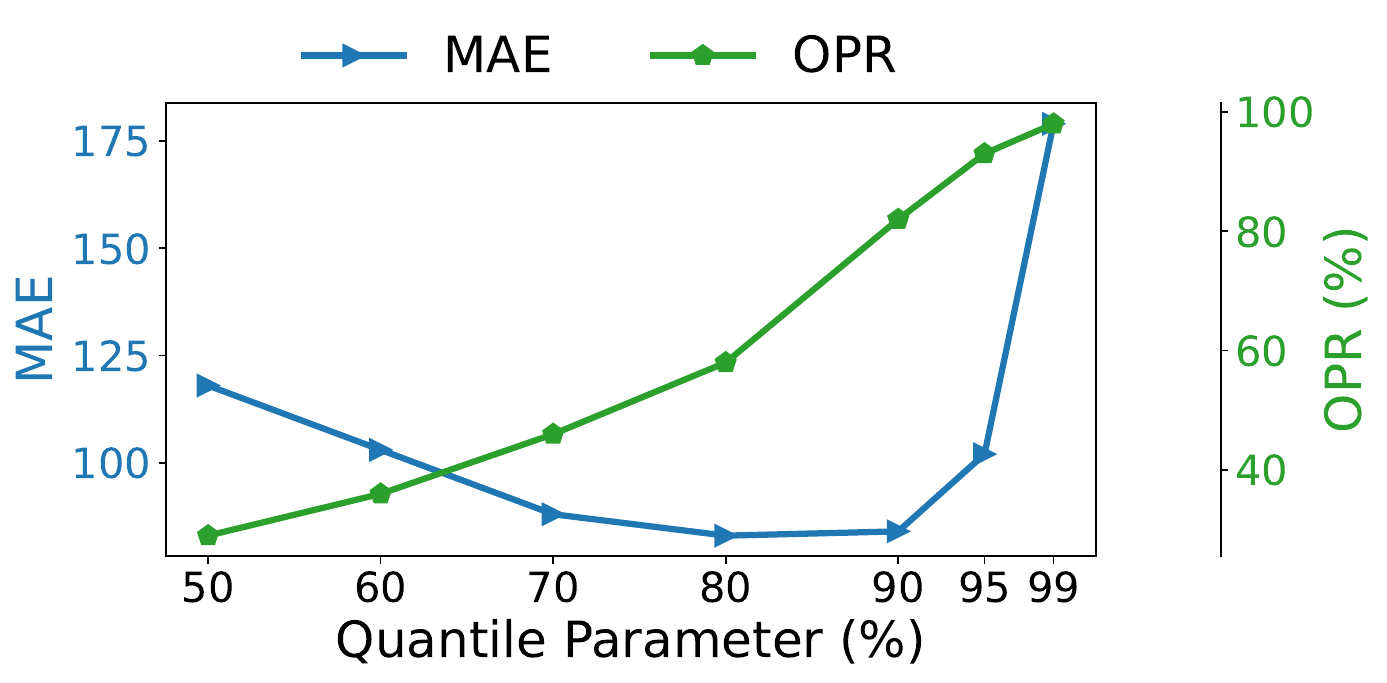} 
	\caption{Sensitivity analysis of quantile parameter $\tau$ on prediction results in MAE and OPR (\cref{sec:res-sen}).}\label{fig:res-sen}
\end{figure}

\subsection{Sensitivity to Quantile Parameter}\label{sec:res-sen}

We evaluate \SYSTEM{}'s sensitivity to the quantile parameter ($\tau$ in \Cref{eq:quantile}) by examining its impact on upgrade outcomes (number of scheduled software upgrades, number of completed software upgrades, and cancellation rate) and prediction metrics (MAE and OPR) for the same UG in~\cref{sec:res-sim}. We sweep $\tau$ across $\{$0.5, 0.6, 0.7, 0.8, 0.9, 0.95, 0.99$\}$ to observe the resulting trends.

In~\Cref{fig:res-sen2}, the number of scheduled upgrades and cancellations initially fluctuate, but both decrease at higher $\tau$, reaching a minimum at 0.95 and 0.99. This is because moderate overpredictions reduce cancellation, but excessive overprediction leads to fewer upgrades being scheduled. Completed upgrades show a more nuanced trend, influenced by both predicted durations and upgrade priorities.

In~\Cref{fig:res-sen}, OPR steadily increases with $\tau$, peaking at 0.99, as higher quantiles lead to more conservative (overpredicted) durations. Interestingly, MAE decreases with $\tau$ initially, and then increases. This is because the data distribution is skewed: the mean (the point that minimizes mean squared error) and median (the point that minimizes MAE) do not align, meaning that a standard regression focused on the average will produce a higher MAE than one focused on specific quantiles. When we explored data, we observed that the data distribution is left-skewed, where the median is higher than the mean. Therefore, a model that overpredicts relative to the mean by using a higher quantile is actually moving closer to the true median, which mathematically reduces MAE.


\section{Related Work}\label{sec:related} 

\textbf{Learning-based behavior prediction.} Machine learning has wide applications in predicting system behaviors such as latency~\cite{belay2014ix}, throughput~\cite{li2020thunderbolt}, and energy consumption~\cite{yuan2003energy}. These predictive models serve as valuable tools in addressing resource management and performance optimization challenges~\cite{ipek2005approach,ipek2006efficiently,deng2017memory,bhatia2019perceptron,garza2019bit,shi2019applying,mao2019learning}. CPR employs linear regression to predict multiprocessor performance~\cite{lee2008cpr}. Paragon uses collaborative filtering to predict quality of service performance in datacenter applications~\cite{delimitrou2013paragon}. CALOREE predicts control parameters for dynamic adaptation through hierarchical Bayesian models~\cite{mishra2018caloree}. Seer uses deep learning to predict performance in microservices~\cite{gan2019leveraging}. These work shares a common goal of prioritizing maximum prediction accuracy, assuming it will directly translate to optimal system outcomes. Ding et al.~\cite{ding2019generative} and NURD~\cite{ding2022nurd}, however, demonstrate that improving prediction accuracy may not lead to improved system outcomes. \SYSTEM{} shares this similar insight but operates in the domain of software upgrades.

\textbf{Prediction-based job scheduling.} Prior work has focused on service jobs scheduling by optimizing resource allocation and execution order to maximize their performance. Many efforts have been made in scheduling service jobs by predicting service job durations~\cite{krishnaswamy2004estimating,curino2014reservation,boutin2014apollo,jalaparti2015network,jyothi2016morpheus,rajan2016perforator,tumanov2016tetrisched,iorgulescu2017don,chung2018stratus,park20183sigma,jajoo2022case}. Corral forecasts job latency by considering future workload characteristics to mitigate the effects of network congestion~\cite{jalaparti2015network}. TetriSched predicts resource requirements for current job execution based on past job executions~\cite{tumanov2016tetrisched}. 3Sigma uses the full distribution of relevant runtime histories to predict job runtimes~\cite{park20183sigma}. SLearn recognizes the input and temporal sensitivity of job runtimes and introduces a sampling technique~\cite{jajoo2022case}. Prior work optimizes prediction accuracy for near-optimal scheduling. In contrast, we introduce a constrained optimization problem in software upgrades, showing accuracy alone is not enough to improve system outcomes.

\textbf{Network change scheduling.} Prior work has studied network changes and scheduling. Janus plans network changes while minimizing the risk by adapting to traffic dynamics~\cite{alipourfard2019risk}. CORNET is a framework for quick and easy adaptation of network change management~\cite{mahimkar2021composition}. Dionysus is a system for fast and consistent network updates in software-defined networks~\cite{jin2014dynamic}. Unlike these systems, \SYSTEM{} targets datacenter-scale software upgrades, tackling their distinct scheduling challenges and operational constraints.

\section{Conclusion} \label{sec:conclusion}

Software upgrades are vital for datacenter reliability but remain under-explored. To address this, we present the first large-scale characterization and analysis using data from Meta's real-world production datacenters. We introduce \SYSTEM{}, a cost-aware duration prediction framework that enhances scheduling throughput and upgrade efficiency while meeting the SLOs. Real-world evaluation shows \SYSTEM{} outperforms existing baseline. We hope this work sparks further research on datacenter-scale upgrades.

\section*{Acknowledgements}
This work was supported by 2021 Meta Research Award on Statistics for Improving Insights, Models, and Decisions.

\bibliography{reference}
\bibliographystyle{mlsys2026}

\appendix

\section{Feature Analysis}\label{sec:feature}

\textbf{Features.} \Cref{tab:upgrade_features} summarizes the 24 features used for upgrade duration prediction. We organize them into three categories: (1) hardware structural features, which capture platform-level resource capacity and architectural characteristics; (2) software state features, which describe the firmware type and version transition involved in the upgrade; and (3) workflow and reset features, which reflect procedural complexity and reset semantics. This categorization reflects the underlying factors that influence upgrade duration: platform-dependent flashing behavior, version-specific migration paths, and reset-induced downtime. Together, these features model both static hardware constraints and dynamic upgrade workflow characteristics, enabling accurate prediction of the highly variable firmware upgrade durations observed in practice.

\textbf{Feature Importances.} We obtain feature importance from the trained Gradient Boosting Tree models used for upgrade duration prediction. Specifically, we extract the built-in importance scores provided by the tree ensemble, which quantify each feature’s contribution to reducing prediction error across all decision splits. These importance values therefore reflect how frequently and effectively a feature is used to partition the data when modeling upgrade duration.

\begin{figure}[htbp]
\centering
\includegraphics[width=\linewidth]{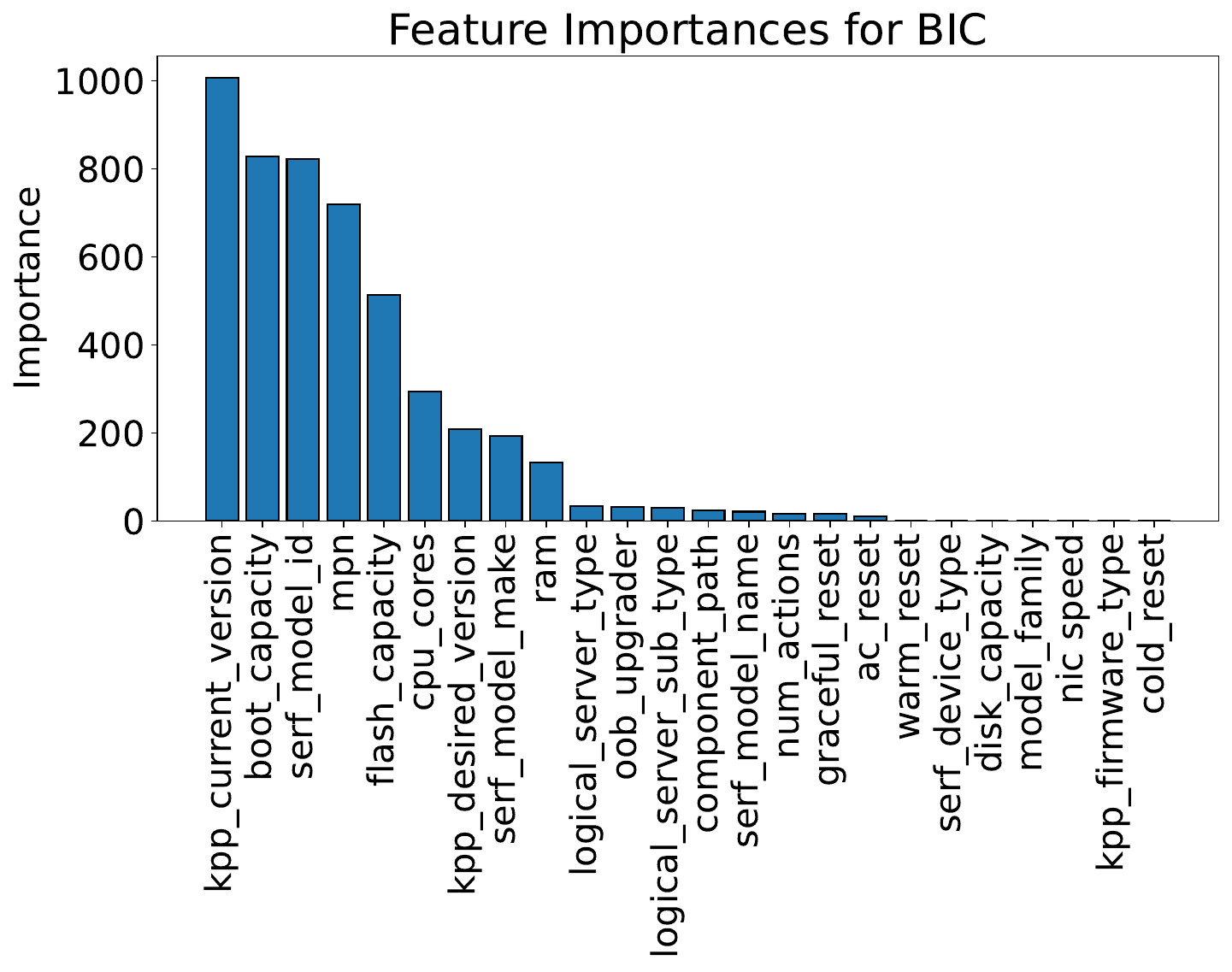}
\caption{Feature importance for training on BIC upgrades.}
\label{fig:fi_bic}
\end{figure}

\begin{figure}[htbp]
\centering
\includegraphics[width=\linewidth]{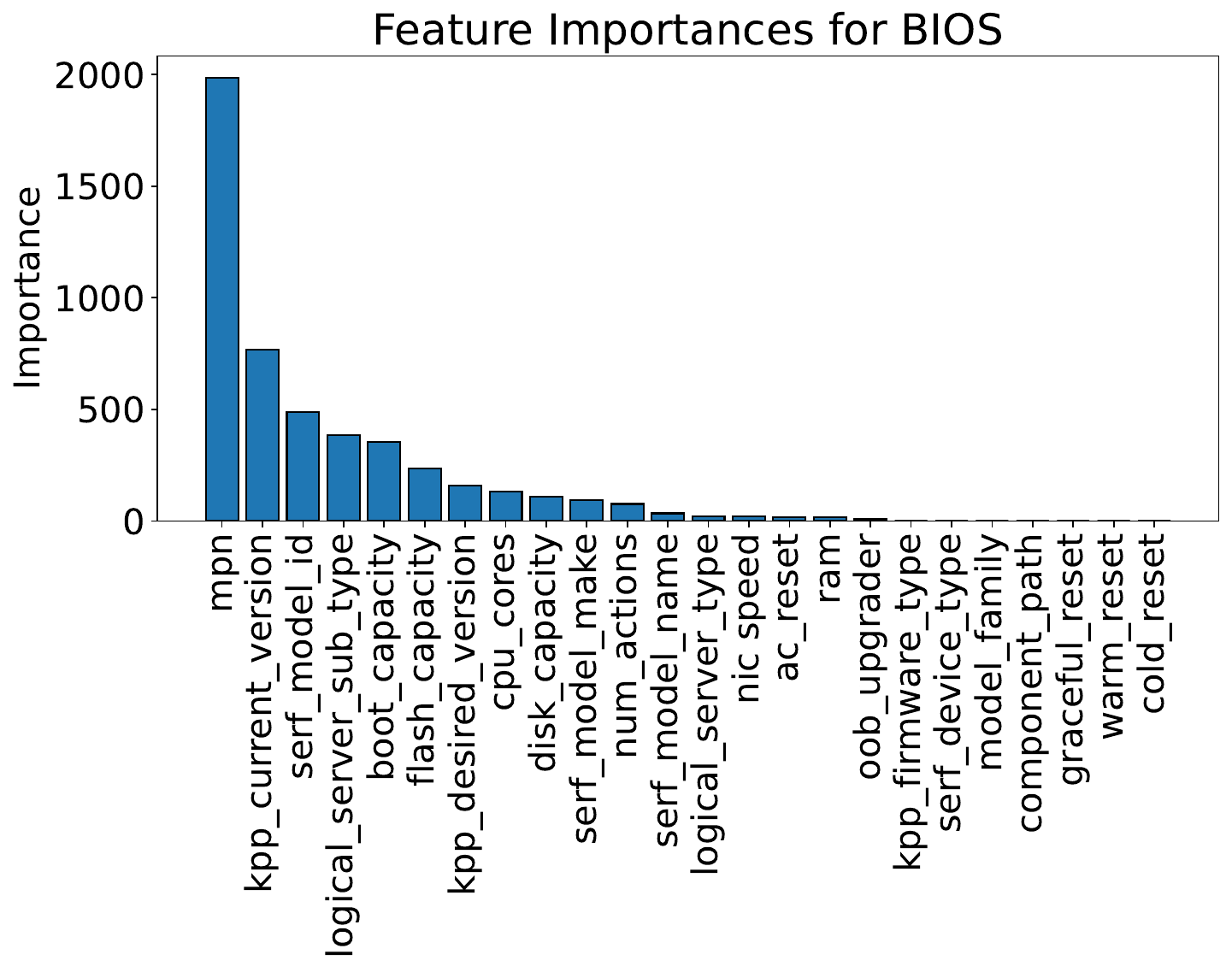}
\caption{Feature importance for training on BIOS upgrades.}
\label{fig:fi_bios}
\end{figure}

\begin{figure}[htbp]
\centering
\includegraphics[width=\linewidth]{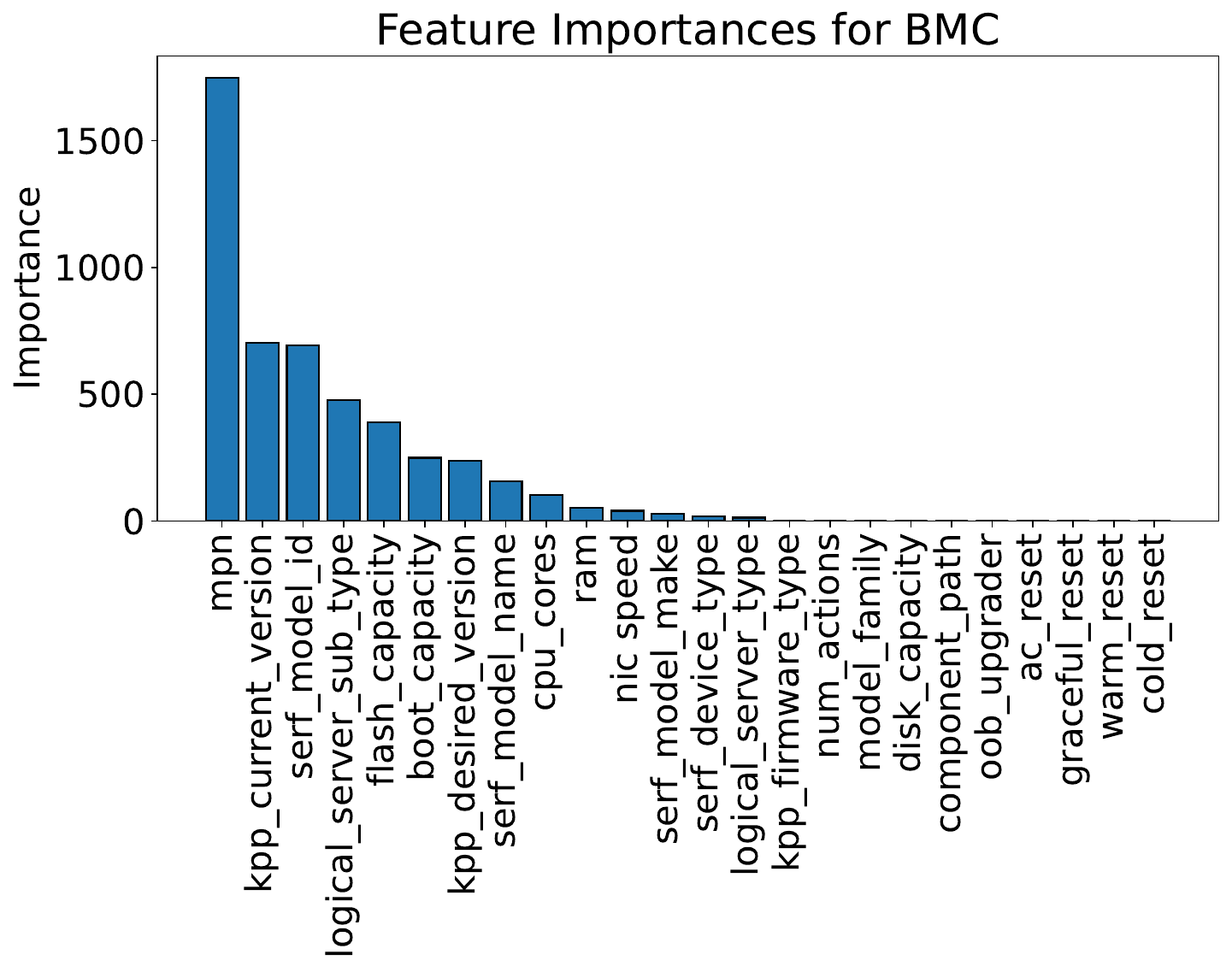}
\caption{Feature importance for training on BMC upgrades.}
\label{fig:fi_bmc}
\end{figure}

\begin{figure}[htbp]
\centering
\includegraphics[width=\linewidth]{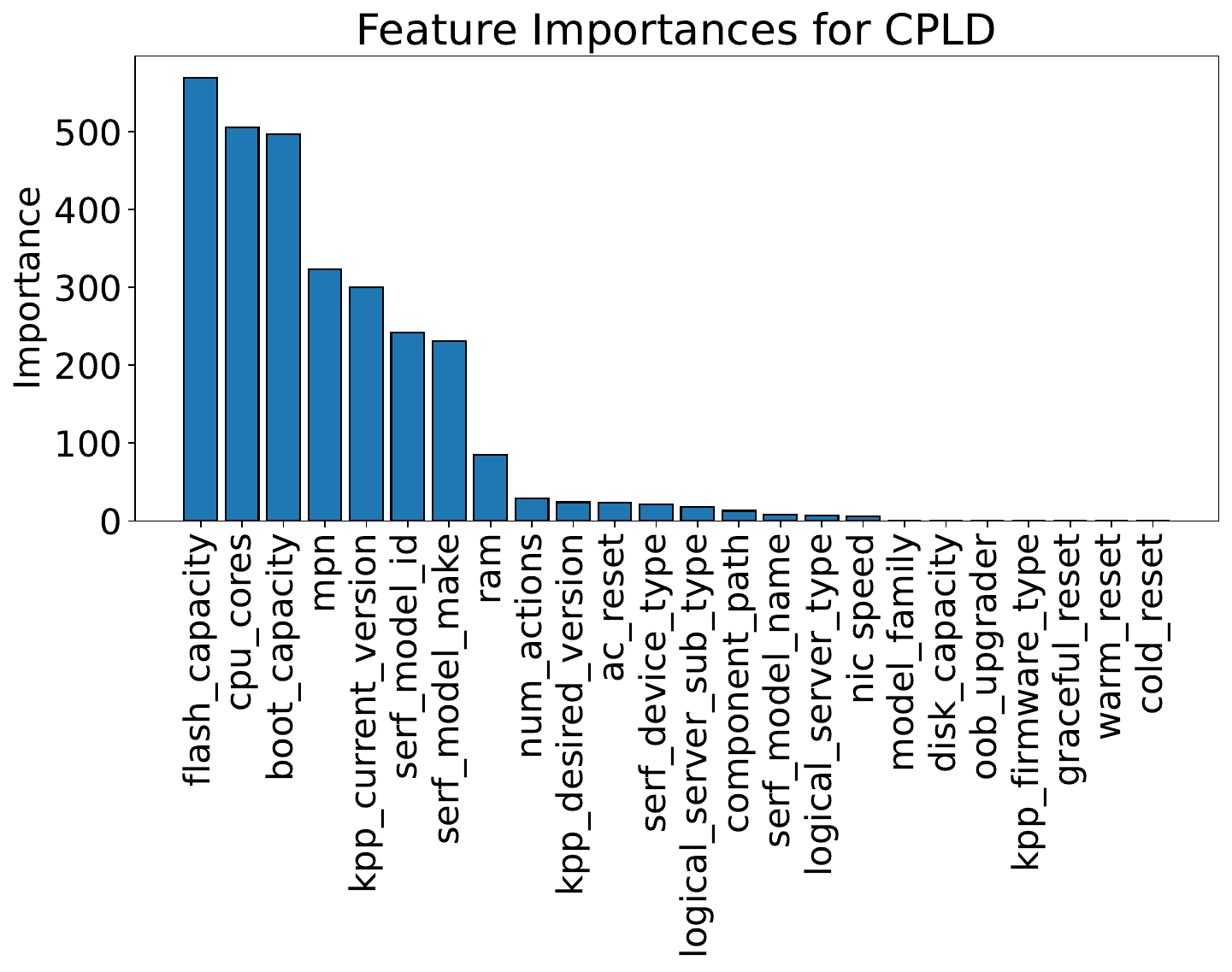}
\caption{Feature importance for training on CPLD upgrades.}
\label{fig:fi_cpld}
\end{figure}

\begin{figure}[htbp]
\centering
\includegraphics[width=\linewidth]{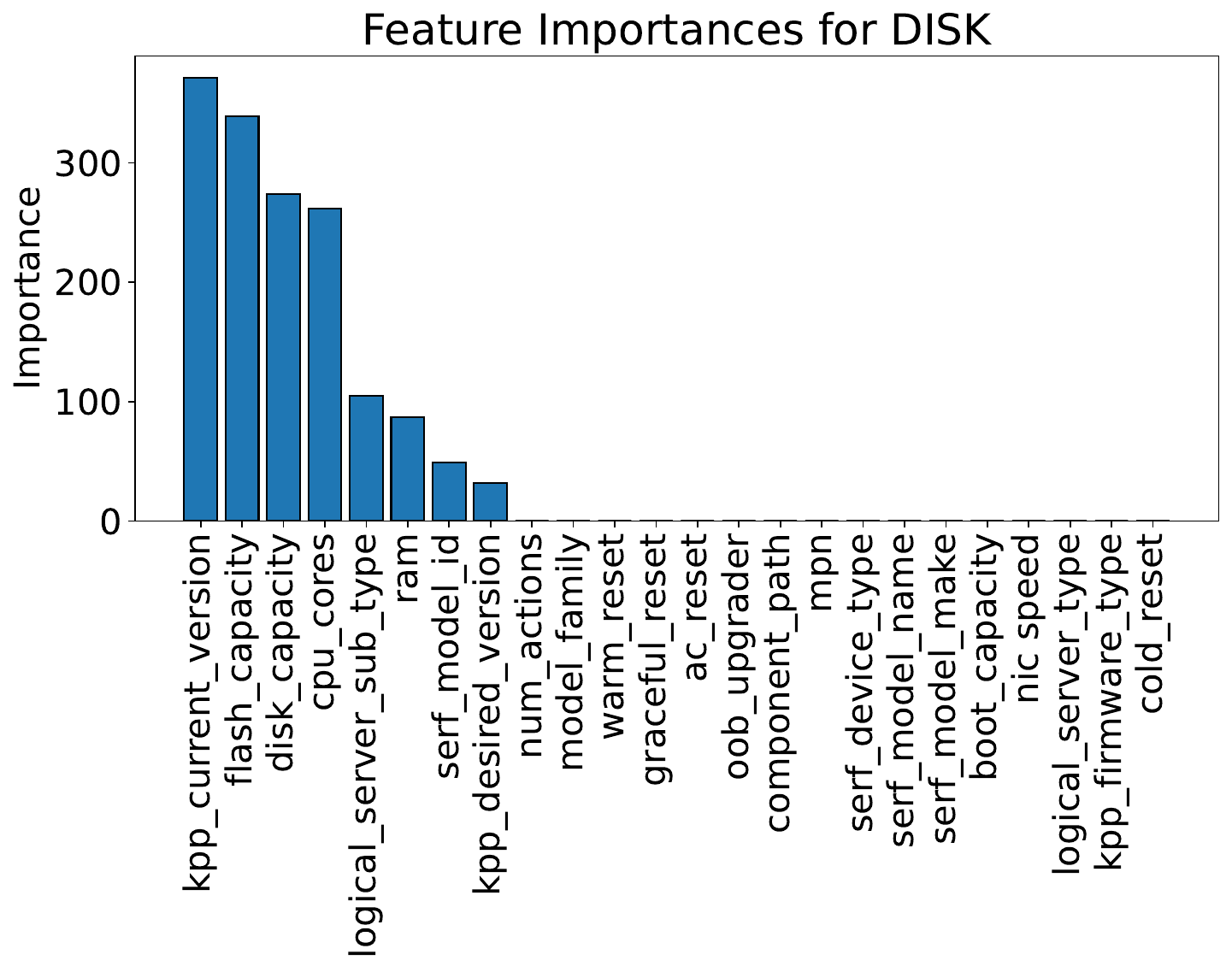}
\caption{Feature importance for training on DISK upgrades.}
\label{fig:fi_disk}
\end{figure}

\begin{figure}[htbp]
\centering
\includegraphics[width=\linewidth]{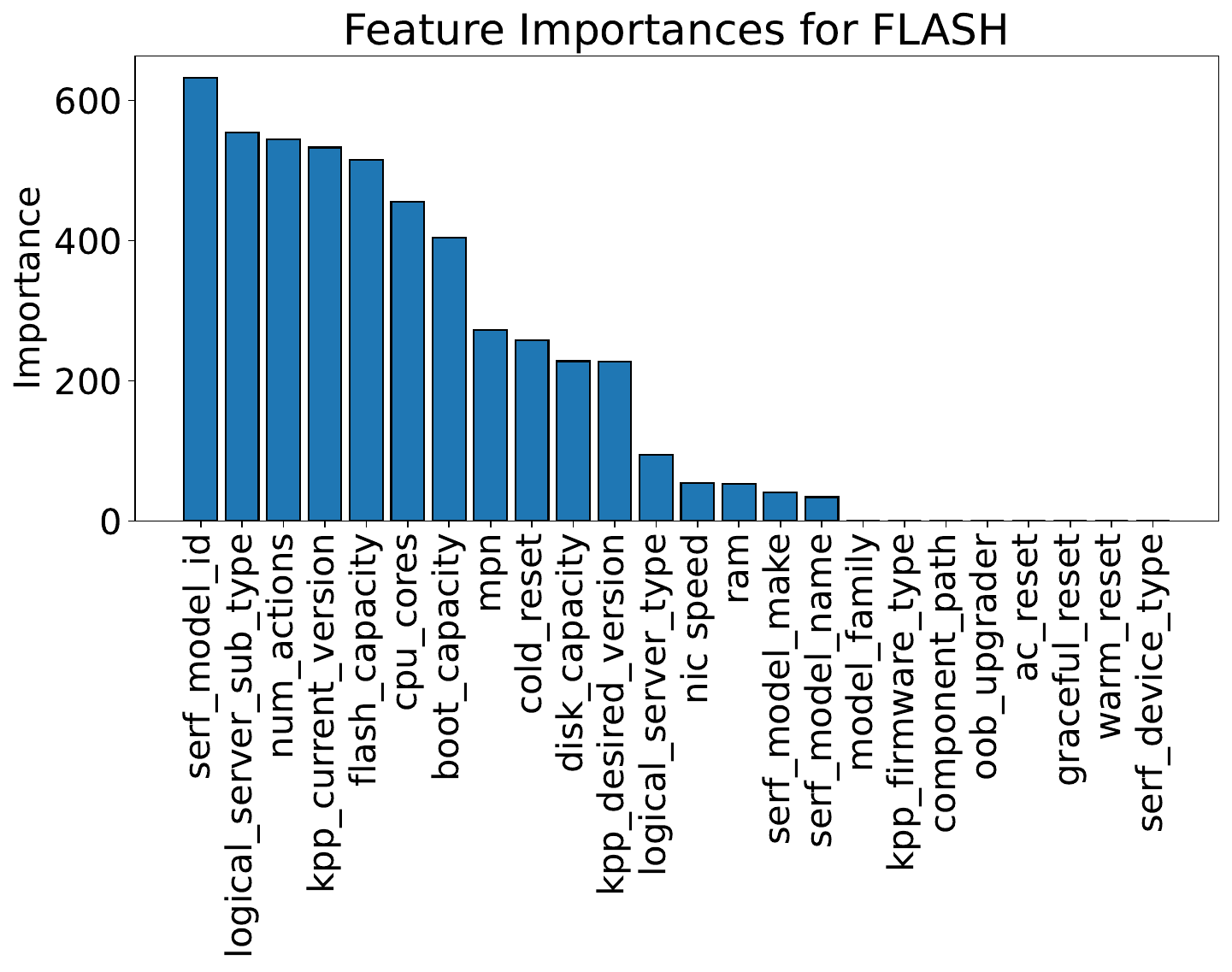}
\caption{Feature importance for training FLASH upgrades.}
\label{fig:fi_flash}
\end{figure}

\begin{figure}[htbp]
\centering
\includegraphics[width=\linewidth]{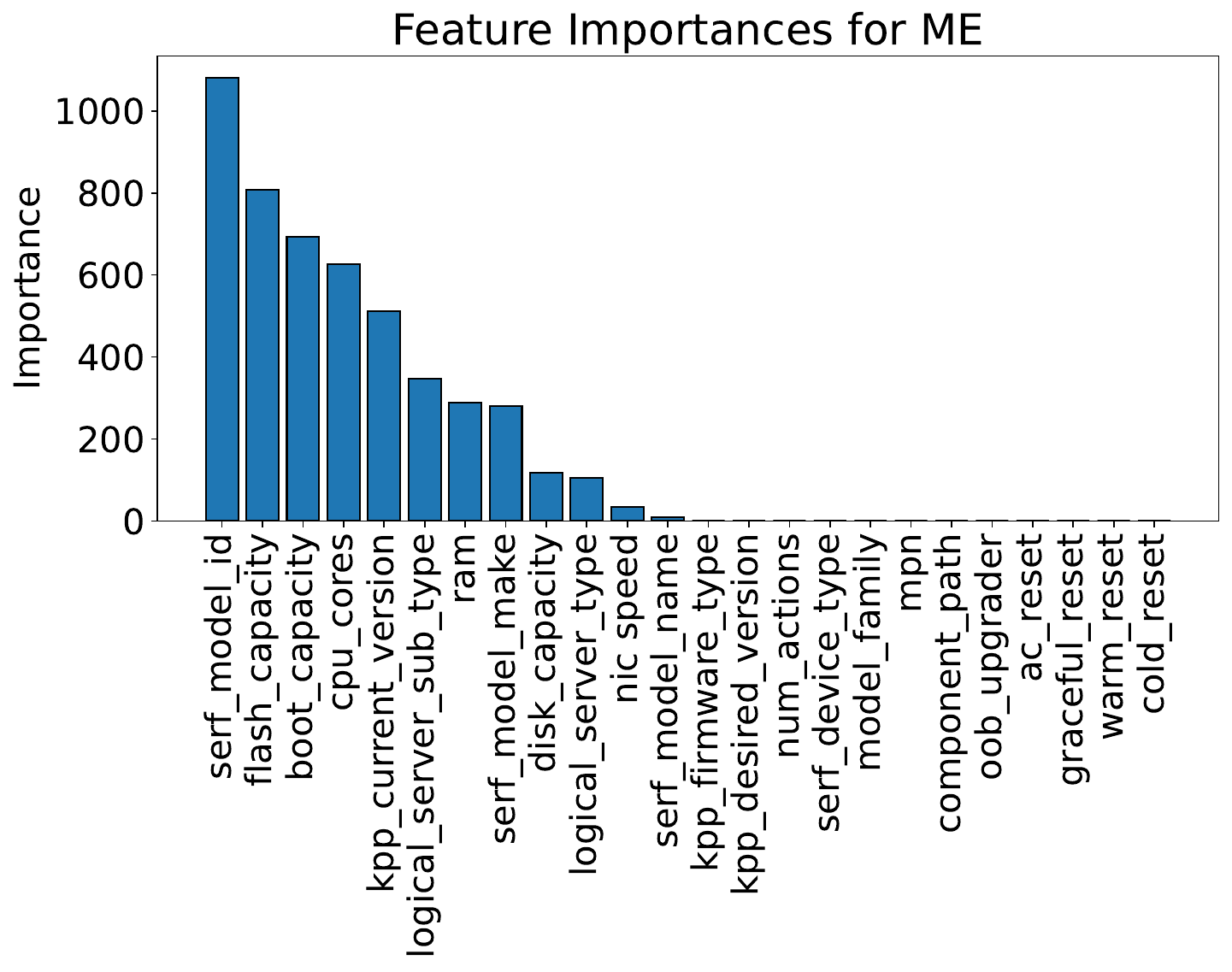}
\caption{Feature importance for training on ME upgrades.}
\label{fig:fi_me}
\end{figure}

\begin{figure}[htbp]
\centering
\includegraphics[width=\linewidth]{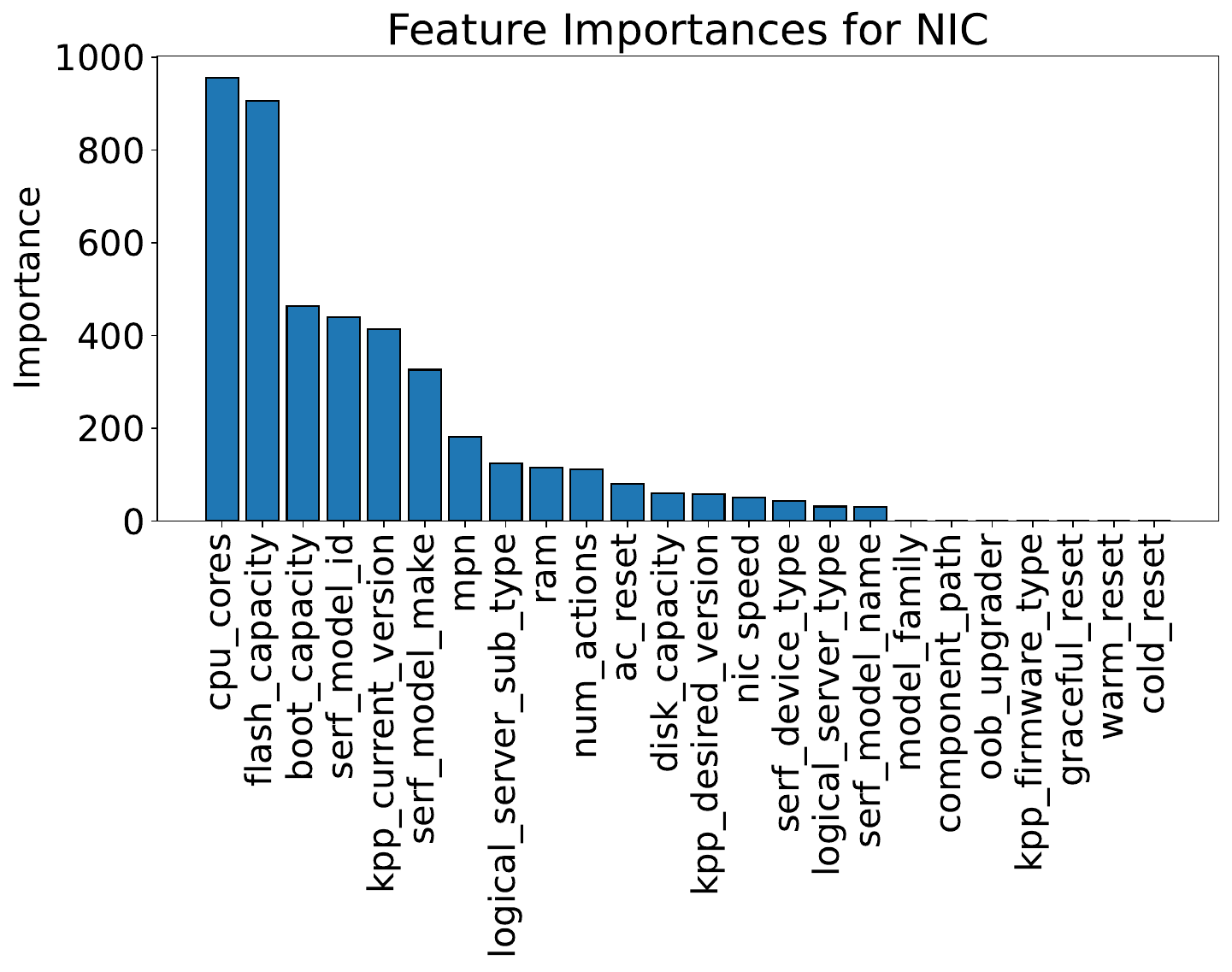}
\caption{Feature importance for training on NIC upgrades.}
\label{fig:fi_nic}
\end{figure}

\Cref{fig:fi_bic,fig:fi_bios,fig:fi_bmc,fig:fi_cpld,fig:fi_disk,fig:fi_flash,fig:fi_me,fig:fi_nic} visualize the feature importances for each firmware upgrade type. First, different firmware types exhibit distinct dominant features, indicating heterogeneous upgrade mechanisms. For example, mpn (manufacturer part number) is the top feature for BIOS and BMC upgrades, suggesting that component-specific revisions are influencing factors in these categories. In contrast, NIC upgrades are dominated by cpu\_cores, while ME upgrades are primarily driven by serf\_model\_id. FLASH upgrades show a more balanced importance distribution across serf\_model\_id, logical\_server\_sub\_type, and num\_actions, indicating stronger interaction between platform identity and procedural complexity. These differences confirm that no single factor universally influences upgrade duration.

Second, version transition features appear among the top predictors across nearly all firmware types. In particular, kpp\_current\_version ranks within the top five for BIC, BIOS, DISK, FLASH, NIC, BMC, and ME. kpp\_desired\_version also appears in several firmware types (e.g., BIC, BIOS, BMC, DISK). This suggests that upgrade duration depends on the version-to-version transition rather than solely on hardware capacity. The current software state influences migration paths, compatibility checks, and post-upgrade validation behavior.

Third, platform identity features emerge as important across firmware types. Features such as serf\_model\_id, logical\_server\_sub\_type, serf\_model\_make, and mpn appear in the top rankings. This indicates that hardware generation, deployment role, and vendor-specific characteristics affect upgrade behavior. Even when different firmware types emphasize different primary drivers, platform descriptors remain structurally influential. Together, these results demonstrate that accurate firmware upgrade duration prediction requires jointly modeling software state transitions and platform-specific characteristics, with firmware-dependent variation in their relative importance.

\begin{table*}[htbp]
\centering
\small
\caption{Features Used for Software/Firmware Upgrade Duration Prediction}
\label{tab:upgrade_features}
\begin{tabular}{p{2.6cm} p{3.2cm} p{2.2cm} p{6.5cm}}
\toprule
\textbf{Feature} & \textbf{Full Name} & \textbf{Category} & \textbf{Description / Relevance to Duration} \\
\midrule

\multicolumn{4}{l}{\textit{Hardware Structural Features}} \\
\midrule

cpu\_cores & Number of CPU Cores & Compute & Total logical/physical cores available. Higher parallelism may accelerate validation, decompression, or initialization phases. \\

ram & Installed Memory (RAM) & Memory & Total system memory. Insufficient memory may slow post-upgrade initialization or trigger swap overhead. \\

disk\_capacity & Total Disk Capacity & Storage & Total installed storage capacity. Larger storage systems may require longer scanning, validation, or metadata checks. \\

flash\_capacity & Flash Storage Capacity & Firmware Storage & Capacity of onboard flash used for firmware images. Larger flash components increase write and verification time. \\

boot\_capacity & Boot Partition Capacity & Boot Device & Size of boot device/partition. Affects bootloader update and firmware flashing duration. \\

nic\_speed & Network Interface Speed & Network & Maximum NIC link speed. Influences image transfer time and orchestration latency. \\

serf\_model\_id & Server Model Identifier & Platform & Internal hardware model identifier capturing platform generation and architectural constraints. \\

serf\_model\_make & Server Manufacturer & Vendor & Hardware vendor. Firmware tooling and flashing behavior vary across manufacturers. \\

serf\_model\_name & Server Model Name & Platform & Specific server product model. Strong determinant of firmware architecture and upgrade process. \\

serf\_device\_type & Device Type & Hardware Type & Physical device category. Influences reset strategy and orchestration workflow. \\

model\_family & Model Family & Hardware Generation & Hardware generation grouping. Captures architectural differences across generations. \\

mpn & Manufacturer Part Number & Component & Exact component part number. Different revisions may exhibit different flashing durations. \\

component\_path & Hardware Component Path & Topology & Logical/physical path to upgraded component. May impact access and validation latency. \\

\midrule
\multicolumn{4}{l}{\textit{Software State Features}} \\
\midrule

kpp\_current\_version & Current Firmware Version & Version State & Existing firmware/software version. Large version gaps may require migration steps or compatibility checks. \\

kpp\_desired\_version & Target Firmware Version & Version State & Intended post-upgrade version. Determines migration path and validation complexity. \\

kpp\_firmware\_type & Firmware Type Name & Component Type & Type of firmware (e.g., BIOS, NIC). Different firmware types have distinct flashing and reset behaviors. \\

logical\_server\_type & Logical Server Type & System Role & High-level server role (e.g., compute, storage). Different roles require different validation workflows. \\

logical\_server\_sub\_type & Logical Server Sub-Type & System Role & More granular role classification (e.g., GPU node, metadata server). Captures specialization effects. \\

\midrule
\multicolumn{4}{l}{\textit{Workflow and Reset Features}} \\
\midrule

num\_actions & Number of Upgrade Actions & Workflow Complexity & Total number of upgrade steps (flash, reboot, verify). Proxy for procedural complexity and duration. \\

oob\_upgrader & Out-of-Band Upgrade Flag & Upgrade Mechanism & Indicates whether upgrade is performed via out-of-band controller (e.g., BMC). May reduce downtime but add orchestration overhead. \\

ac\_reset & AC Power Reset Required & Reset Type & Indicates full AC power cycle requirement. Typically longest reset duration. \\

graceful\_reset & Graceful Software Reset & Reset Type & Controlled OS-level restart. Usually shorter and more predictable. \\

warm\_reset & Warm Reboot & Reset Type & Hardware reset without full power cycle.  \\

cold\_reset & Cold Reboot & Reset Type & Full reboot including hardware reinitialization without full AC cycle. \\
\bottomrule
\end{tabular}
\end{table*}

\section{Comparing to Other Learning Models}\label{sec:models}

\begin{figure*}[htbp]
\centering
\includegraphics[width=\linewidth]{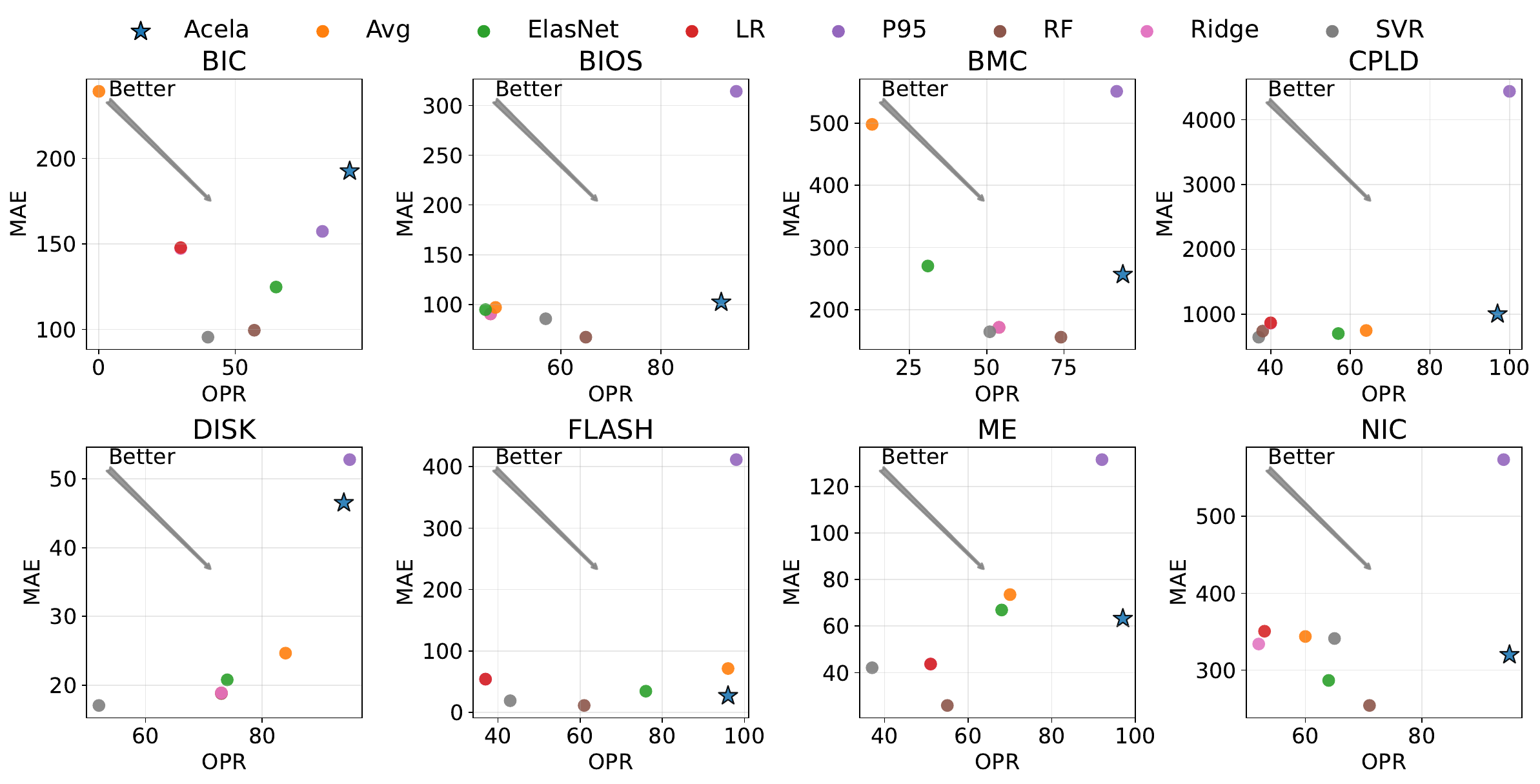}
\caption{Tradeoffs between OPR and MAE across firmware types.}
\label{fig:app-tradeoff}
\end{figure*}

To understand the learning model choices of \SYSTEM{}, we compare it against seven representative baselines spanning both learning-based and heuristic approaches. These baselines reflect common modeling strategies used for duration prediction, including symmetric-loss regression models and simple heuristics:
\begin{itemize}
    \item \textbf{P95}: A heuristic baseline that predicts durations using the 95th percentile of historical data, reflecting conservative worst-case provisioning commonly used in production systems.
    \item \textbf{RF (Random Forest)}: An ensemble tree-based model that improves robustness through bagging and non-linear feature interactions.
    \item \textbf{ElasNet (Elastic Net)}: A linear model combining L1 and L2 regularization to balance sparsity and stability.
    \item \textbf{SVR (Support Vector Regression)}: A kernel-based model that captures non-linear relationships through margin-based optimization.
    \item \textbf{Ridge}: A linear regression model with L2 regularization to reduce variance.
    \item \textbf{LR (Linear Regression)}: A standard least-squares regression model that predicts the conditional mean.
    \item \textbf{Avg}: A heuristic baseline that predicts durations using the average per firmware type.
\end{itemize}

These baselines primarily optimize prediction accuracy under symmetric loss functions, which aim to minimize average error but do not account for the asymmetric operational consequences of prediction errors in upgrade scheduling.

To evaluate model performance, we consider two complementary metrics: Mean Absolute Error (MAE) and Overprediction Rate (OPR). MAE captures prediction accuracy, while OPR measures the fraction of predictions that overestimate actual durations and directly reflects the ability to meet the system's SLO (i.e., ensuring upgrades complete within the upgrade window). As discussed in the paper, these two objectives are inherently misaligned: minimizing MAE encourages predicting the conditional mean, while meeting the SLO requires intentional overprediction to avoid underestimation risks.

\Cref{fig:app-tradeoff} shows this MAE–OPR tradeoff across firmware types. Models optimized purely for accuracy (e.g., RF, SVR, Ridge, LR) achieve low MAE but consistently exhibit insufficient OPR, failing to satisfy the SLO requirement. This behavior stems from their reliance on symmetric loss functions, which treat under- and over-predictions equally and therefore center predictions around the conditional mean. As a result, they underpredict too frequently in a system where underprediction carries higher cost.

In contrast, heuristic approaches such as P95 and Avg achieve higher OPR by construction but suffer from large MAE due to their inability to capture fine-grained variability across upgrade types and system contexts. These methods resemble the existing worst-case scheduling strategy and lead to inefficient resource utilization.

\SYSTEM{} achieves the best tradeoff between MAE and OPR. By explicitly modeling asymmetric misprediction costs through quantile regression and selecting models using a cost-aware scoring function, \SYSTEM{} shifts predictions toward slight overestimation while avoiding excessive conservatism. This enables \SYSTEM{} to achieve OPR close to the SLO target (95\%) while maintaining competitive MAE.

Overall, these results highlight a key insight: prediction accuracy alone is insufficient for system optimization. Models that minimize MAE do not necessarily lead to better scheduling outcomes. Instead, \SYSTEM{}'s cost-aware design enables it to align prediction behavior with system-level objectives, achieving a superior balance between accuracy and operational reliability across firmware types.



\end{document}